\documentclass{aa}
\usepackage[varg]{txfonts}
\usepackage{color}

\begin{document}

\title{Detection of 40-48 GHz dust continuum linear polarization towards the Class 0 young stellar object IRAS 16293-2422
  \\ 
  } 

\titlerunning{JVLA Q-band polarization towards IRAS 16293-2422}

\author{Hauyu Baobab Liu\inst{1} 
   \and Yasuhiro Hasegawa\inst{2}
   \and Tao-Chung Ching\inst{3} 
   \and Shih-Ping Lai\inst{4}
   \and Naomi Hirano\inst{5}
   \and Ramprasad Rao\inst{5} 
       } 


\institute{European Southern Observatory (ESO), Karl-Schwarzschild-Str. 2, D-85748 Garching, Germany  
  \and
  Jet Propulsion Laboratory, California Institute of Technology, Pasadena, CA, 91109, USA 
  \and
  National Astronomical Observatories, Chinese Academy of Sciences, China
  \and 
  Institute of Astronomy and Department of Physics, National Tsing Hua University, Hsinchu, Taiwan
  \and 
  Academia Sinica Institute of Astronomy and Astrophysics, P.O. Box 23-141, Taipei, 106 Taiwan
  } 

\date{Received 10 January 2018 / Accepted 05 May 2018}

\abstract {} 
{The aims of this work are to test the feasibility of observing dust polarization at frequency lower than 50 GHz, which is the optically thinner part of the modified black body spectrum; and to clarify whether or not the polarization mechanism is identical or similar to that for (sub)millimeter observations.} 
{We performed the new Karl G. Jansky Very Large Array (JVLA) full polarization observations at 40-48 GHz (6.3-7.5 mm) towards the nearby ($d$ $=$147$\pm$3.4 pc) Class 0 young stellar object (YSO) IRAS 16293-2422, and compare with the previous Submillimeter Array (SMA) observations reported by Rao et al. (2009; 2014). We observed the quasar J1407+2827 which is weakly polarized and can be used as a leakage term calibrator for $<$9 GHz observations, to gauge the potential residual polarization leakage after calibration.} 
{We did not detect Stokes Q, U, and V intensities from the observations of J1407+2827, and constrain (3-$\sigma$) the residual polarization leakage after calibration to be $\lesssim$0.3\%. Limited by thermal noise, we only detect linear polarization from one of the two binary components of our target source, IRAS\,16293-2422\,B. The measured polarization percentages range from less than one percent to few tens of percent. The derived polarization position angles from our observations are in excellent agreement with those detected from the previous observations of the SMA, implying that on the spatial scale we are probing ($\sim$50-1000 au), the physical mechanisms for polarizing the continuum emission do not vary significantly over the wavelength range of $\sim$0.88-7.5 mm.} 
{We hypothesize that the observed polarization position angles trace the magnetic field which converges from large scale to an approximately face-on rotating accretion flow. In this scenario, magnetic field is predominantly poloidal on $>$100 au scales, and becomes toroidal on smaller scales. However, this interpretation remains uncertain due to the high dust optical depths at the central region of IRAS\,16293-2422\,B and the uncertain temperature profile. We suggest that dust polarization at wavelengths comparable or longer than 7\,mm may still trace interstellar magnetic field. Future sensitive observations of dust polarization in the fully optically thin regime will have paramount importance for unambiguously resolving the magnetic field configuration.} 

\maketitle

\begin{figure*}
   \hspace{-0.5cm}
   \begin{tabular}{ p{8.8cm} p{8.8cm} }
     \includegraphics[width=9.5cm]{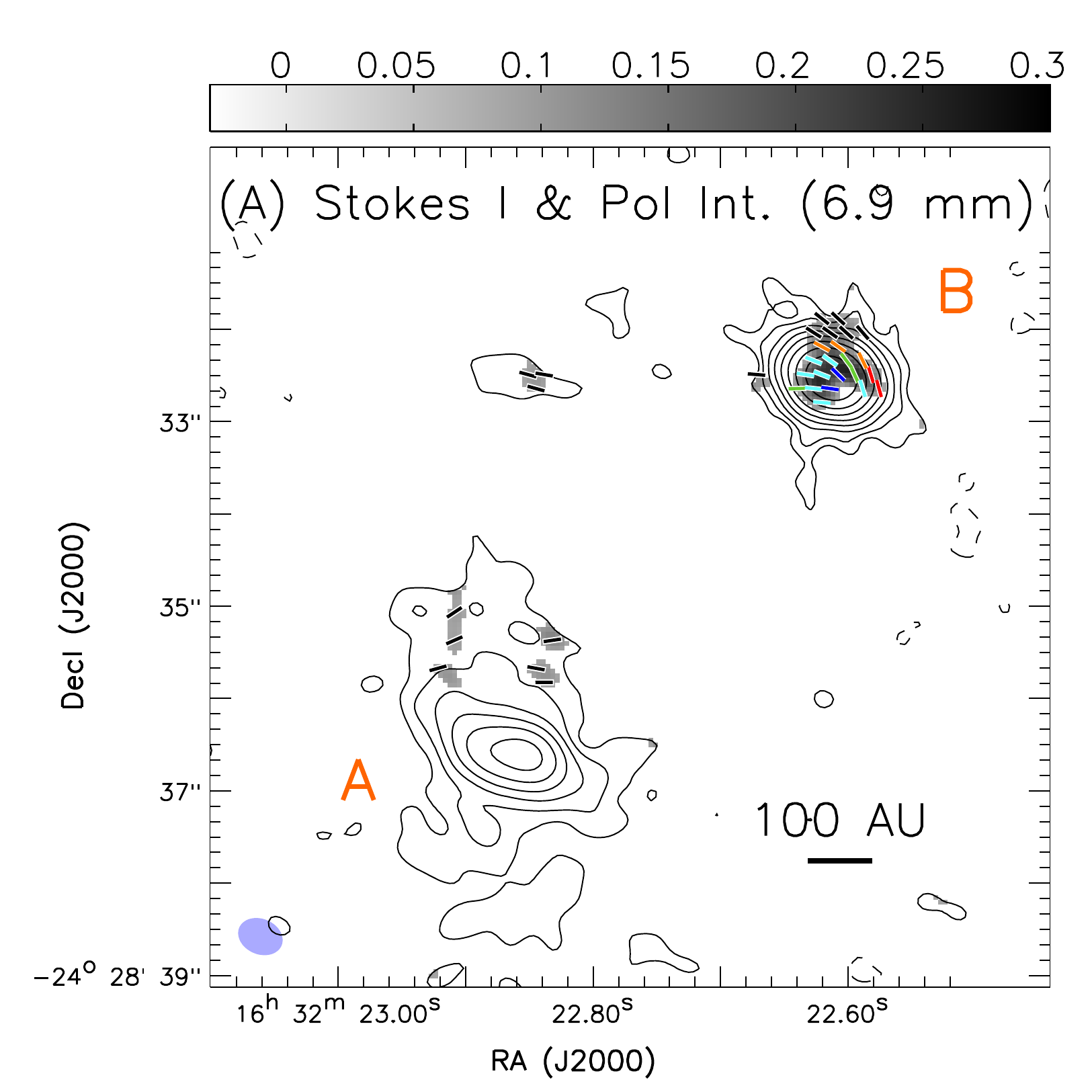} &
     \includegraphics[width=9.5cm]{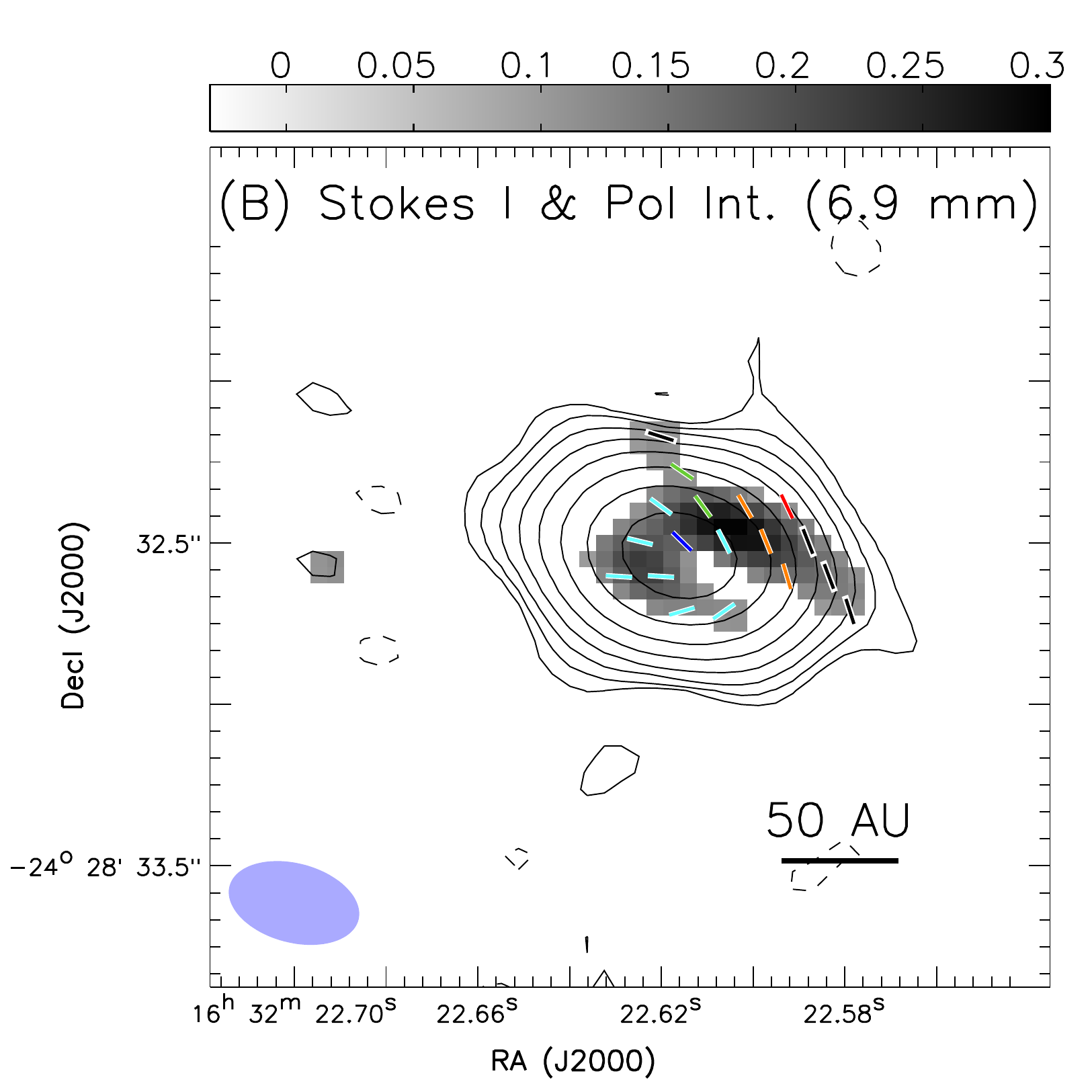} \\
   \end{tabular}
   \caption{\footnotesize{
   JVLA 40-48 GHz continuum images of IRAS\,16293-2422. Panels (A) and (B) show the Robust =2 weighted images [$\theta_{\mbox{\tiny{maj}}}$   $\times$ $\theta_{\mbox{\tiny{min}}}$ $=$ 0$\farcs$49 $\times$ 0$\farcs$38 (P.A. = 66$^{\circ}$)] and the Robust = 0 weighted images [$\theta_{\mbox{\tiny{maj}}}$   $\times$ $\theta_{\mbox{\tiny{min}}}$ $=$ 0$\farcs$39 $\times$ 0$\farcs$24 (P.A. = 74$^{\circ}$)], respectively. Panel B only shows a small area around IRAS\,16293-2422\,B since the Robust = 0 weighted images only detect polarized intensity at $>$3-$\sigma$ from this region. The synthesized beams are shown in bottom left of these panels. Contours present the Stokes I intensities. The polarized intensities are presented in grayscale. Contours in panels (A) and (B) are 0.03 mJy\,beam$^{-1}$ (1$\sigma$, $\sim$0.10 K) $\times$ [-3, 3, 6, 12, 24, 48, 96, 192, 384] and 0.035 mJy\,beam$^{-1}$ (1$\sigma$, $\sim$0.24 K) $\times$ [-3, 3, 6, 12, 24, 48, 96, 192, 384], respectively.
   Contours of negative intensity are dashed. Color bars are in mJy\,beam$^{-1}$ units. The line segments present the measured polarization (E-field) position angles: blue, cyan, green, orange, red, and black colors represent the $<$1\%, 1\%-2\%, 2\%-3\%, 3\%-5\%, 5\%-10\%, and $>$10\% polarization percentages, respectively. The polarization percentages and polarization position angles measured at the centroids of these line segments, are given in Tables \ref{tab:pol_rob2} and \ref{tab:pol_rob0}. 
   }}
   \label{fig:jvla}
\end{figure*}

\section{Introduction}\label{sec:introduction}
Imaging linear polarization of dust continuum is becoming a mature technique for probing  interstellar magnetic field configuration (e.g., Hildebrand et al. 2000) and dust properties (e.g., Kataoka et al. 2015; Yang et al. 2016; Stephens et al. 2017; Hull et al. 2018).
The previous high angular resolution  polarimetric surveys of dust continuum emission towards star-forming regions were mostly carried out at 230-690 GHz bands ( $\sim$0.43-1.3 mm). 
This is to take the advantage of the bright dust emission at these bands, and the not too poor atmospheric transmission for the ground based observatories with large apertures (e.g., Matthews et al. 2009; Dotson et al. 2010; Hull et al. 2014; Zhang et al. 2014).
However, for sources which remain embedded, such as most of the OB star-forming cores, and the Class\,0 and I low-mass young stellar objects (YSOs), dust emission may be highly optically thick at these frequency bands.
In fact, instruments prior to the existence of the Atacama Large Millimeter Array (ALMA) were likely limited by sensitivity, such that the detection of linear  polarization was biased towards the optically thicker and brighter sources. 
It remains important to observe dust polarization at optically thinner, lower frequency bands, to penetrate inside the obscured sources and thereby unambiguously resolve how magnetic field is connected from circumstellar cores to rotating disks, and how dust properties (e.g., size, asymmetry) evolve in these regions.
For regions where the maximum grain size is above a few millimeter due to dust grain growth (e.g., Vorobyov et al. 2018), and if large dust grains can still be efficiently aligned with magnetic field (e.g., Hoang \& Lazarian 2016), polarimetric continuum observations at $\gtrsim$3\,mm bands may trace local magnetic field configuration without being serious confused by foreground and background which do not emit efficiently at long wavelengths due to small grain sizes.

To our knowledge, presently there were only two attempts for constraining linear polarization of dust continuum at frequency lower than 50 GHz (Cox et al. 2015; Liu et al. 2016).
Both of them targeted on the deeply embedded Class 0 YSO NGC\,1333\,IRAS4A.
The angular resolution of the 29 GHz and 37 GHz observations reported by Cox et al. (2015) was very high, and the detection of polarization was limited to a region with rather high brightness temperature, where dust emission may be optically thick.
They stressed that the detected polarization percentages were strikingly high.
The 40-48 GHz observations of Liu et al. (2016) verified that the polarization position angles on $\sim$500-1000 au scales agree with those detected from the Submillimeter Array (SMA) observations at 345 GHz.
Within the central $\sim$50 au radius of NGC\,1333\,IRAS4A1, it became rather difficult to ignore the difference between the polarization position angles observed at 40-48 GHz and 345 GHz, although part of the difference can be attributed to residual polarization leakage (up to $\lesssim$1\%) of those 40-48 GHz observations.
Moreover, the observed polarization position angles in this region at 40-40 GHz and 345 GHz are both offset from what was resolved by Cox et al. (2015) at 29 and 37 GHz, while the reason behind is not yet clear at least to us.

To expand our experience on the observations of dust polarization at $<$90 GHz, we have performed the $\sim$0$\farcs$3 angular resolution polarization observations at 40-48 GHz\footnote{Note that observations at the frequency range of 50-80 GHz are challenging for ground based observatories due to low atmospheric transmission.} towards another nearby ($d$ $\sim$147 pc; Ortiz-Le{\'o}n et al. 2017) Class 0 YSO, IRAS\,16293-2422, using the NRAO Karl G. Jansky Very Large Array (JVLA).
The details of our observations and data reduction are outlined in Section \ref{sec:observation}.
Our results are presented in Section \ref{sec:results}.
Our interpretation for the present observations is discussed in Section \ref{sec:discussion}.
A brief conclusion is given in Section \ref{sec:conclusion}.
We provide our measurements of Stokes I, Q, U intensities at certain sampling positions, in Appendix \ref{appendix:poltable}.
More comparison of our 40-48 GHz observations with the 341.5 GHz observations of SMA is shown in Appendix \ref{appendix:taper}.
In Appendix \ref{appendix:line}, we base on our interpretation in Section \ref{sec:discussion} to outline a radiative transfer effect which can anomalously polarize spectra lines.

\begin{figure*}
   \vspace{-1cm}
   \begin{tabular}{ p{5.6cm} p{5.6cm} p{5.6cm}  }
     \includegraphics[width=6.2cm]{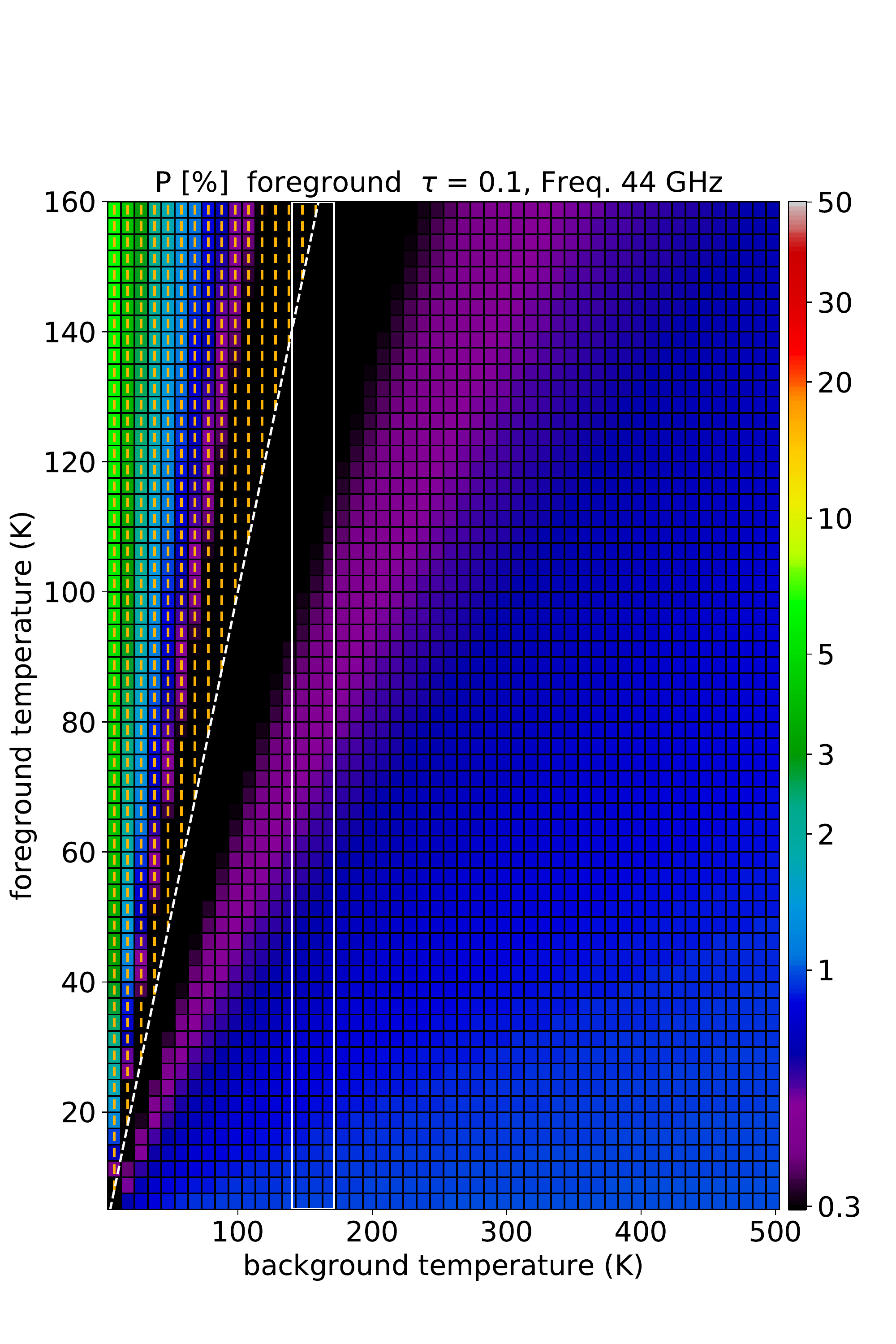} &
     \includegraphics[width=6.2cm]{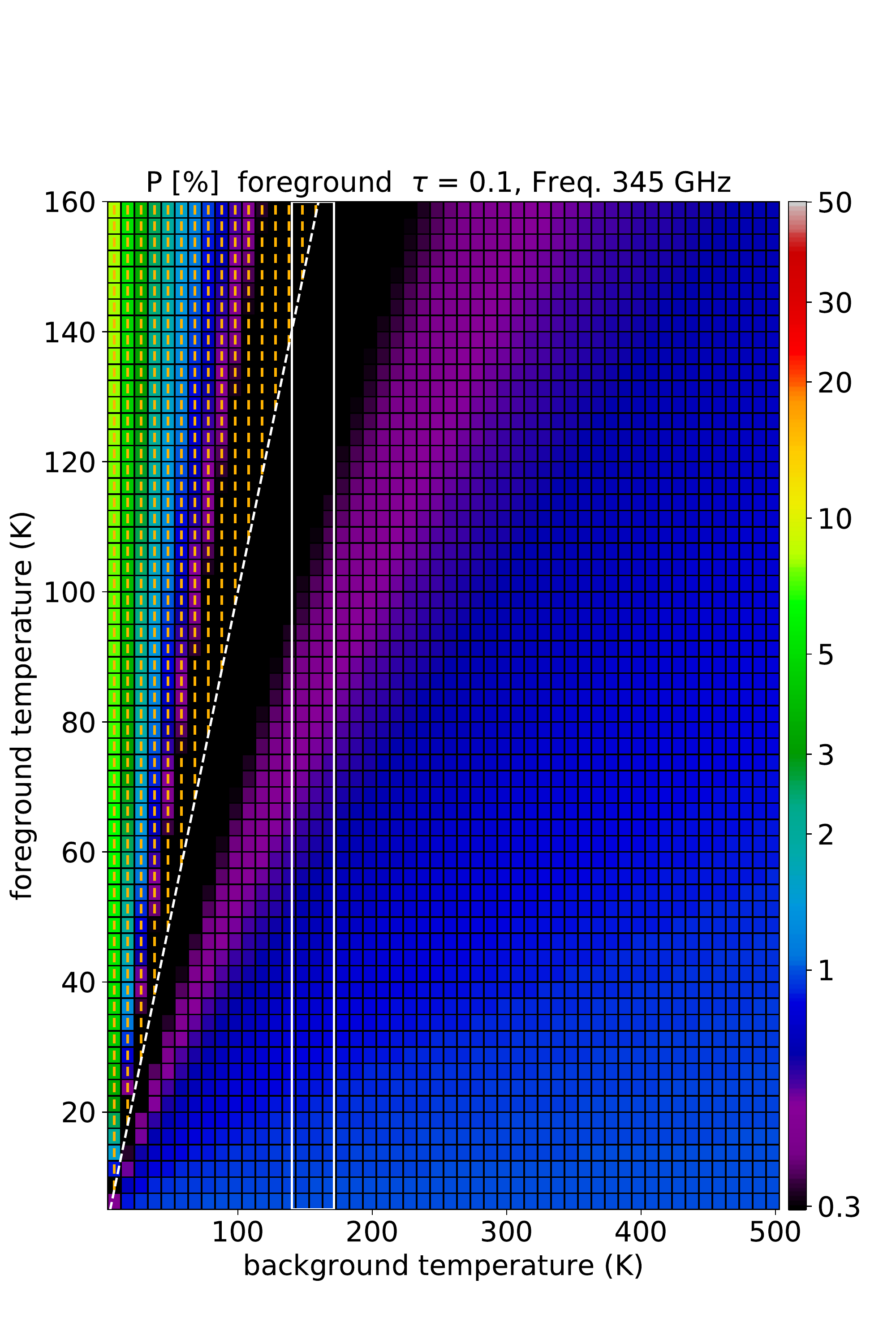} &
     \includegraphics[width=6.2cm]{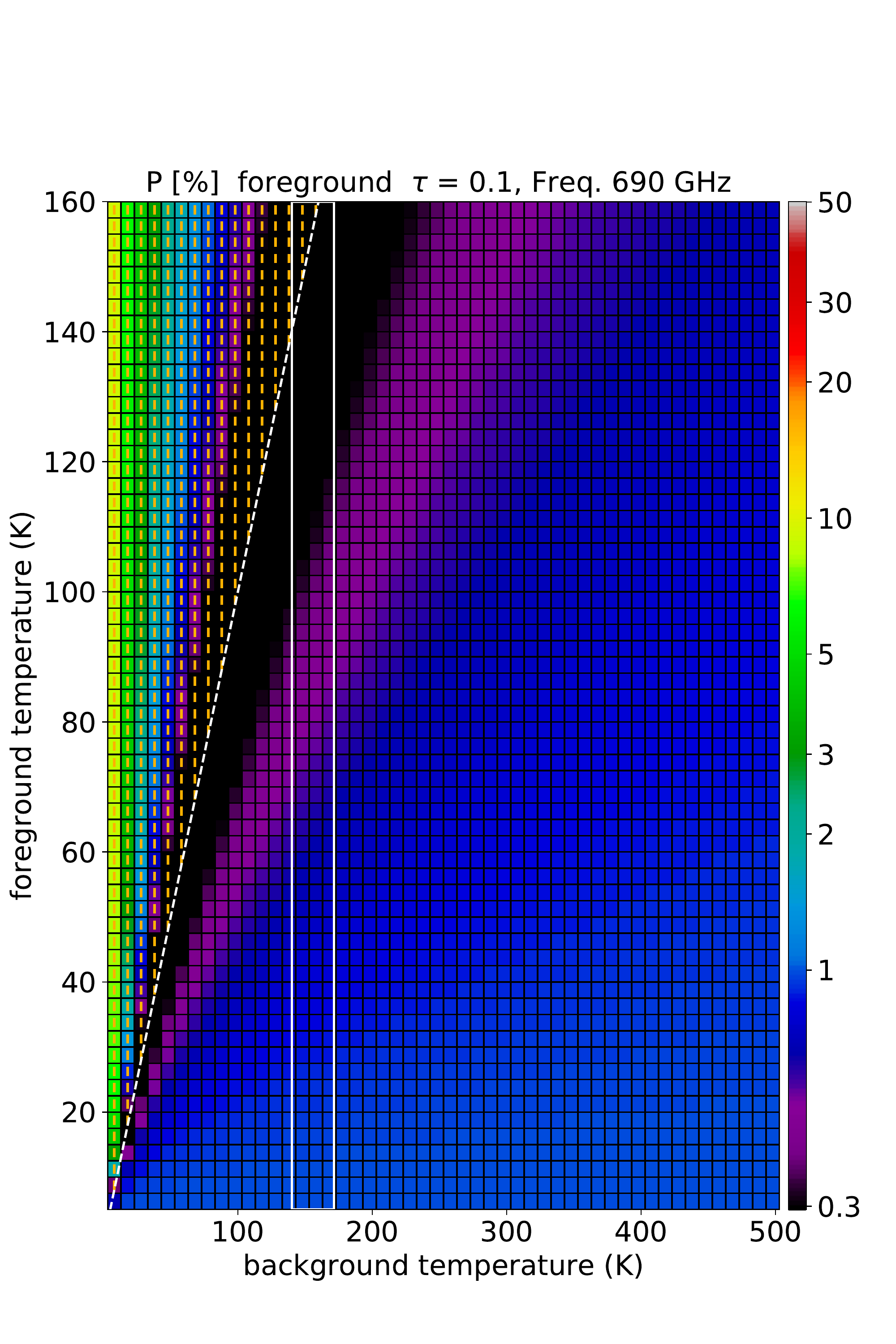}  \\
   \end{tabular}
   
   \vspace{-1.5cm}
   \begin{tabular}{ p{5.6cm} p{5.6cm} p{5.6cm}  }
     \includegraphics[width=6.2cm]{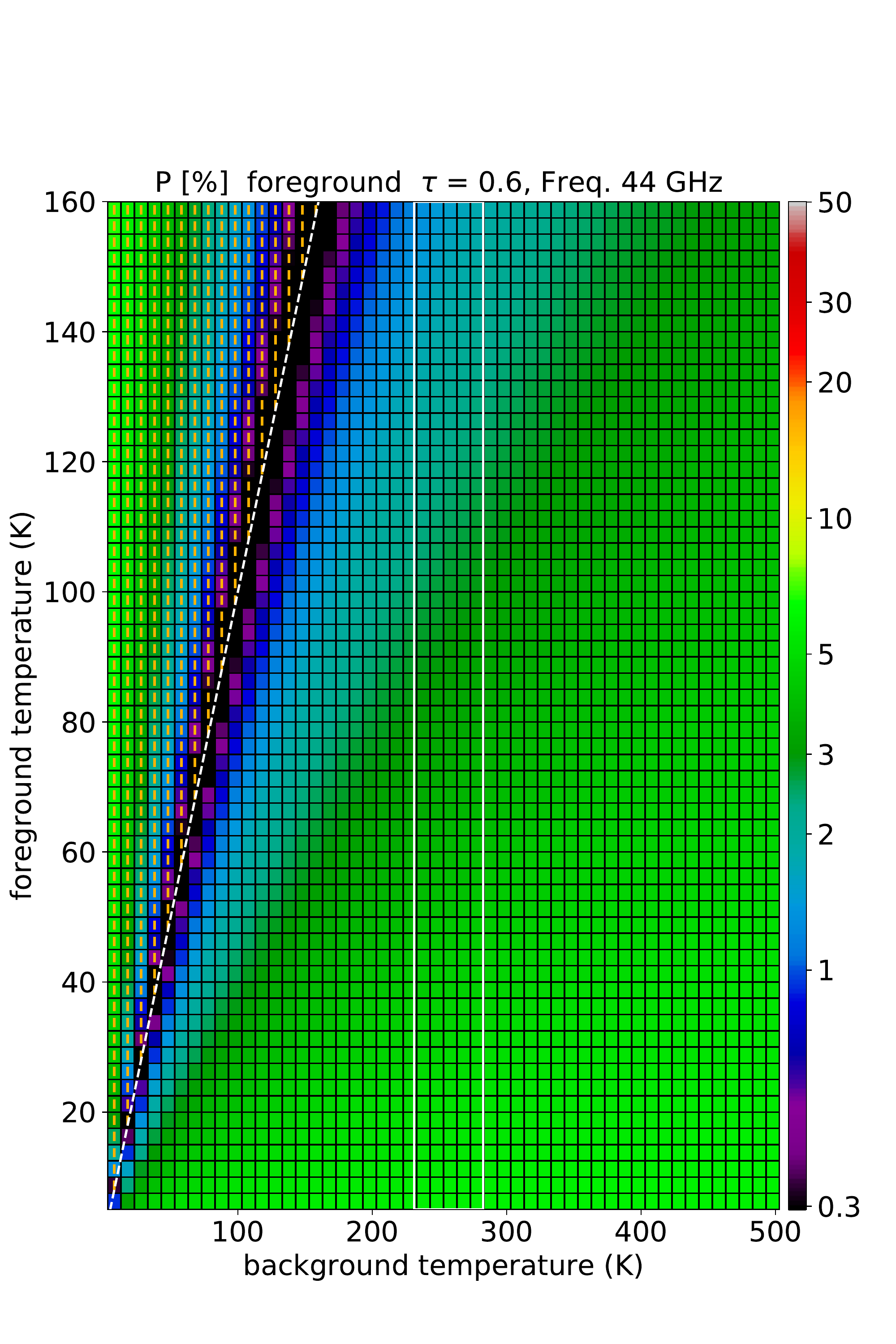} &
     \includegraphics[width=6.2cm]{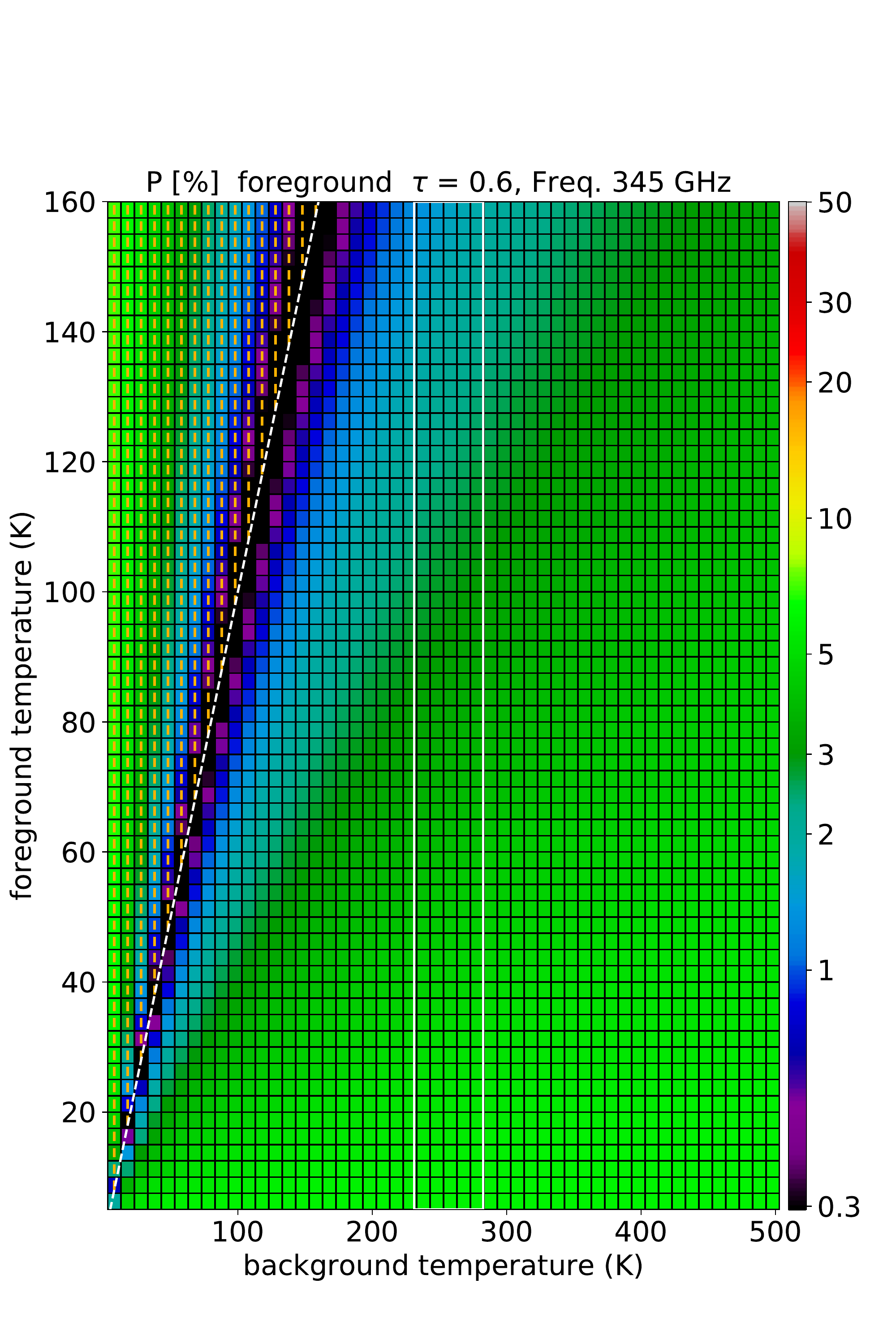} &
     \includegraphics[width=6.2cm]{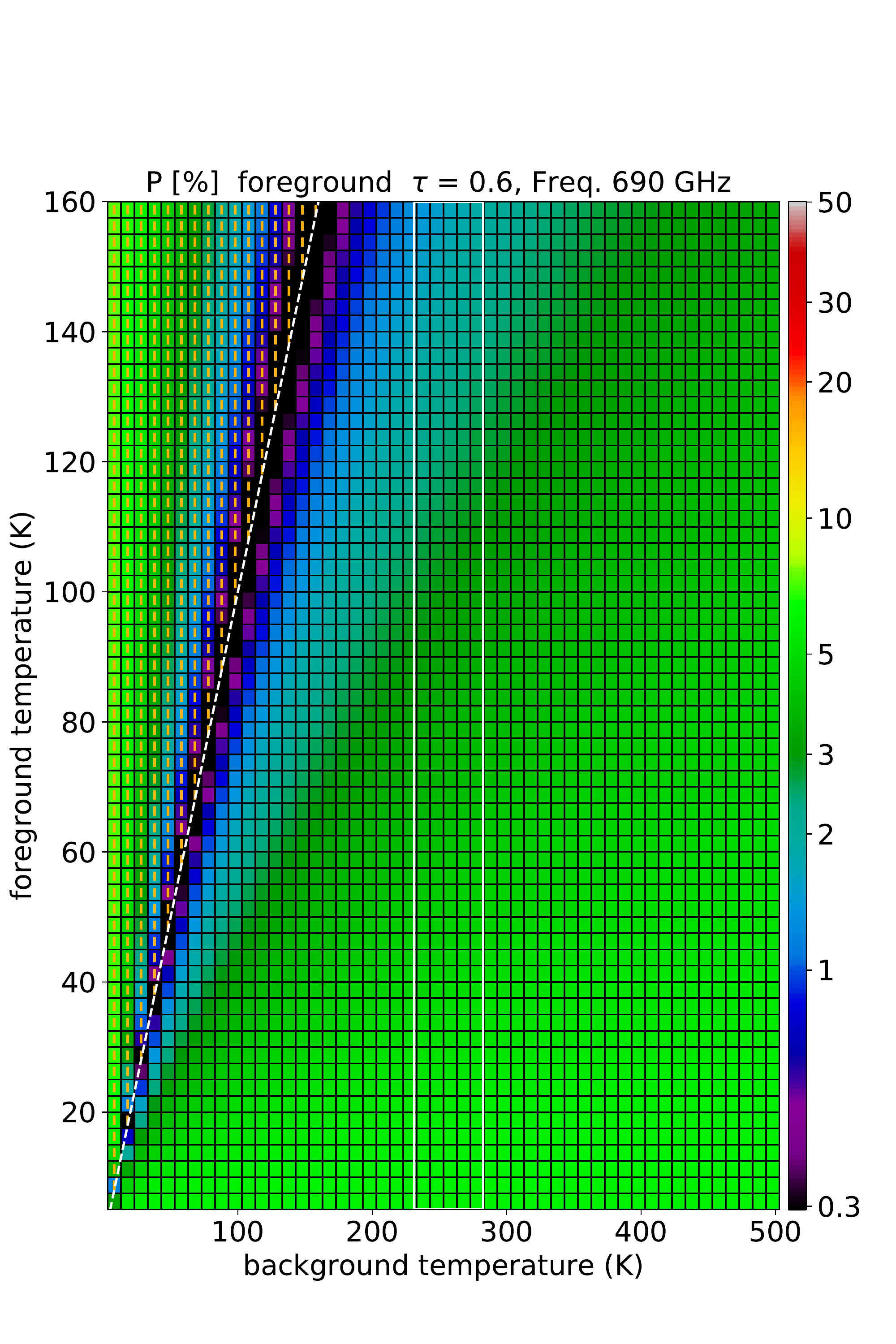}  \\
   \end{tabular}  
   \vspace{-0.3cm}
   \caption{
      \footnotesize{
      Derived polarization percentage (color) from the two-component model introduced in Section \ref{sub:mechanism} (Equation \ref{eqn:multicomponent}), with $\alpha=10\%$ (Equation \ref{eqn:alpha}), and observing frequency $\nu=$44 (left column), 345 (middle column), and 690 GHz (right column). Color bars are in \% units. Horizontal and vertical axes are dust temperatures of the background and foreground components $T_{fg}$ and $T_{bg}$; the resolutions of $T_{fg}$ and $T_{bg}$ in our calculations are the same with the horizontal and vertical grids. White dashed lines show $T_{fg}=T_{bg}$. Top and bottom rows show the cases of $\tau=$0.1 and 0.6, respectively. The orange line segments indicate that the observed polarization position angle is identical to that of the emission of foreground component; otherwise the observed polarization position angle is 90$^{\circ}$ offset from that of the emission of foreground component. Vertical white lines show the extinction corrected peak brightness temperature of IRAS\,16293-2422\,B (141$\pm$14\,K)$/e^{-\tau}$.
      }
   }
   \label{fig:Permodel}
\end{figure*}

\begin{figure*}
   \vspace{-1cm}
   \begin{tabular}{ p{5.6cm} p{5.6cm} p{5.6cm}  }
     \includegraphics[width=6.2cm]{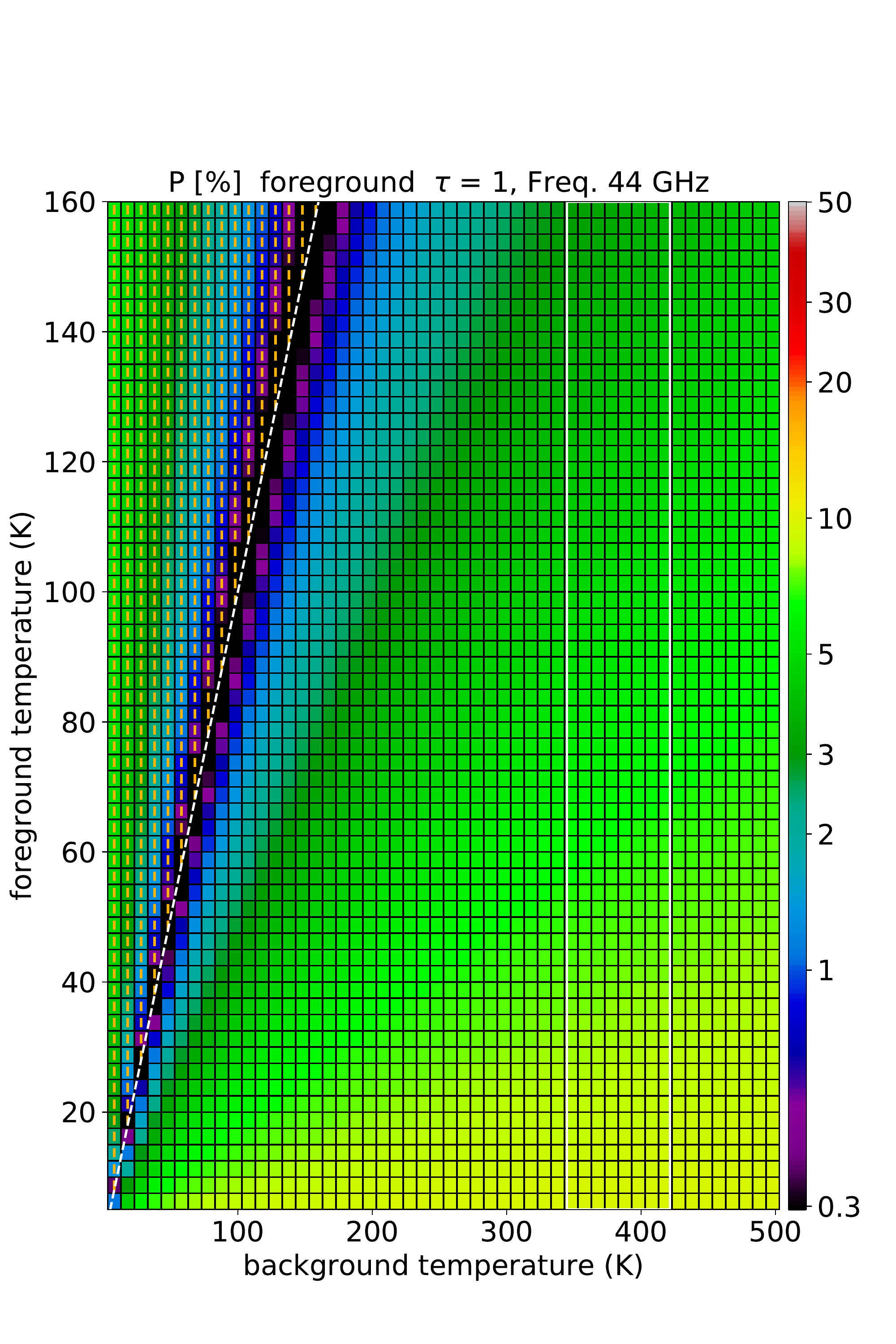} &
     \includegraphics[width=6.2cm]{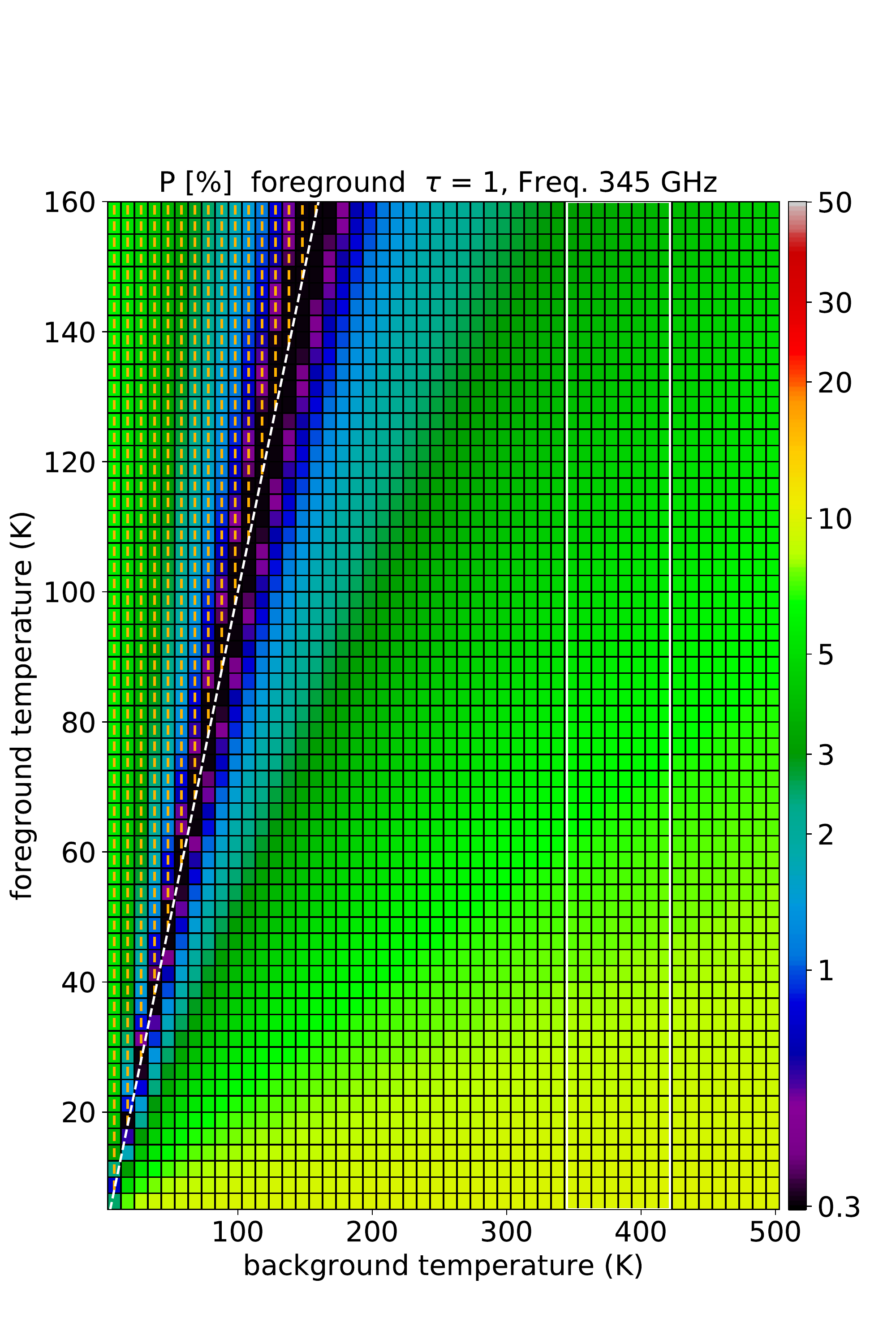} &
     \includegraphics[width=6.2cm]{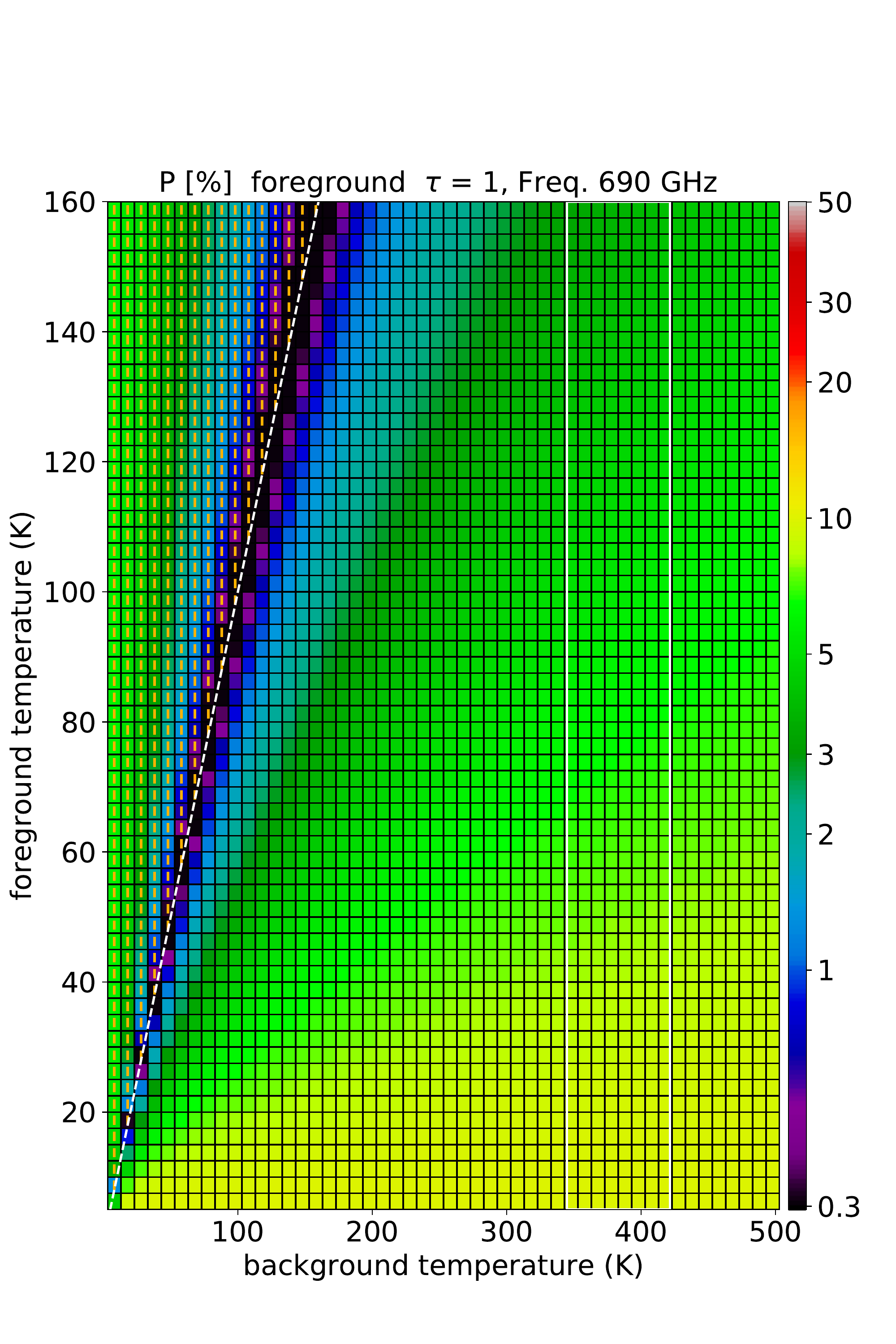}  \\
   \end{tabular}
   
   \vspace{-1.5cm}
   \begin{tabular}{ p{5.6cm} p{5.6cm} p{5.6cm}  }
     \includegraphics[width=6.2cm]{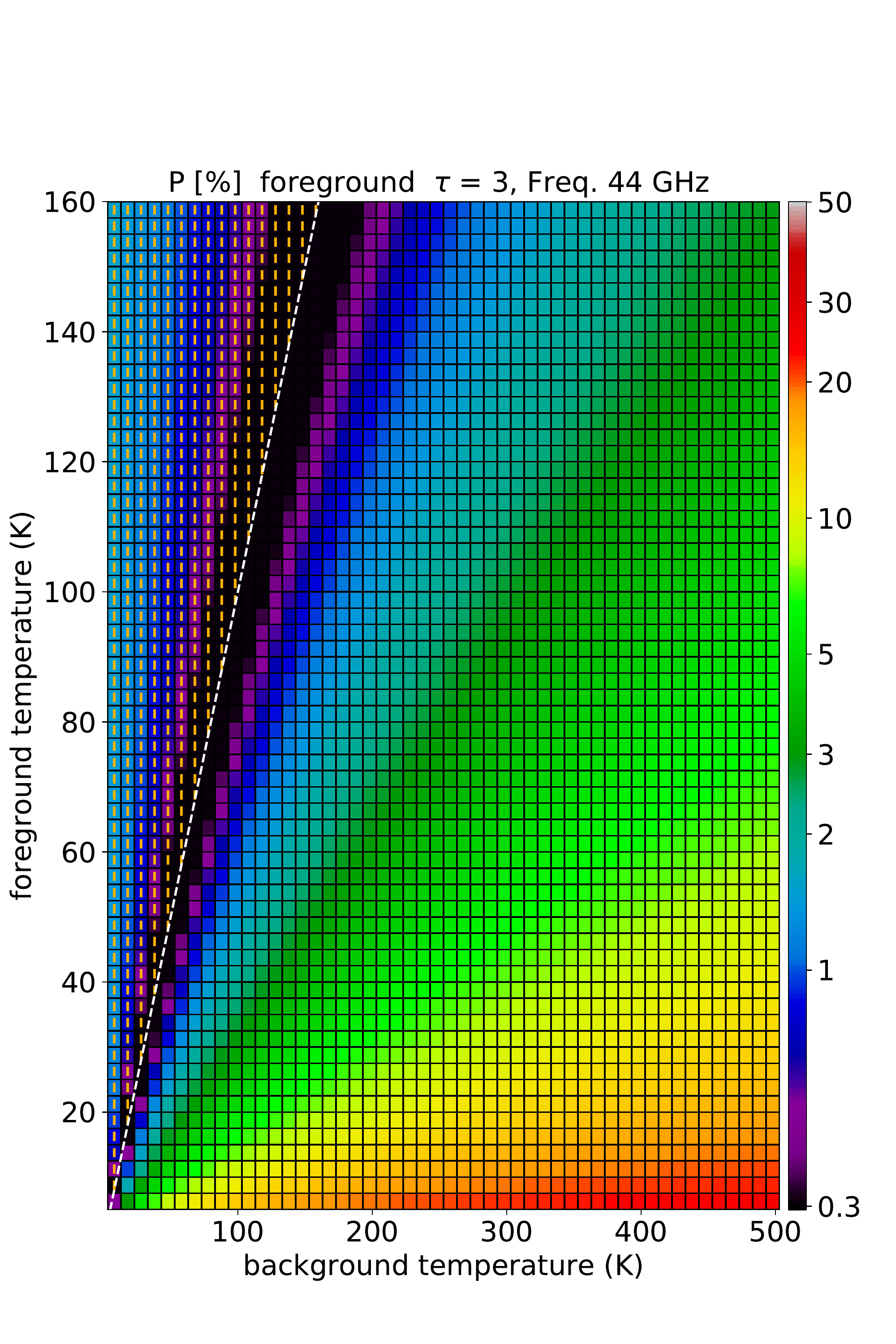} &
     \includegraphics[width=6.2cm]{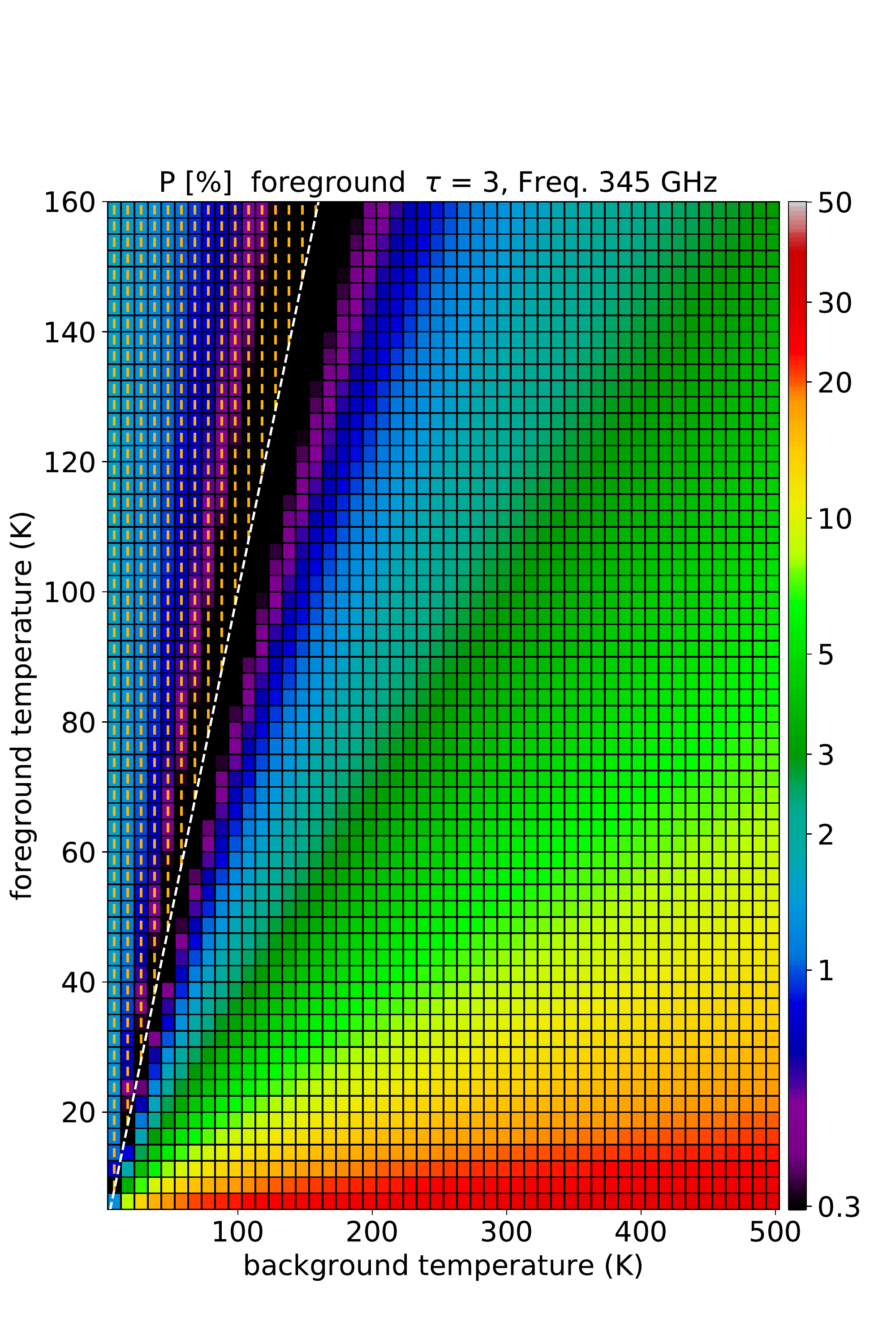} &
     \includegraphics[width=6.2cm]{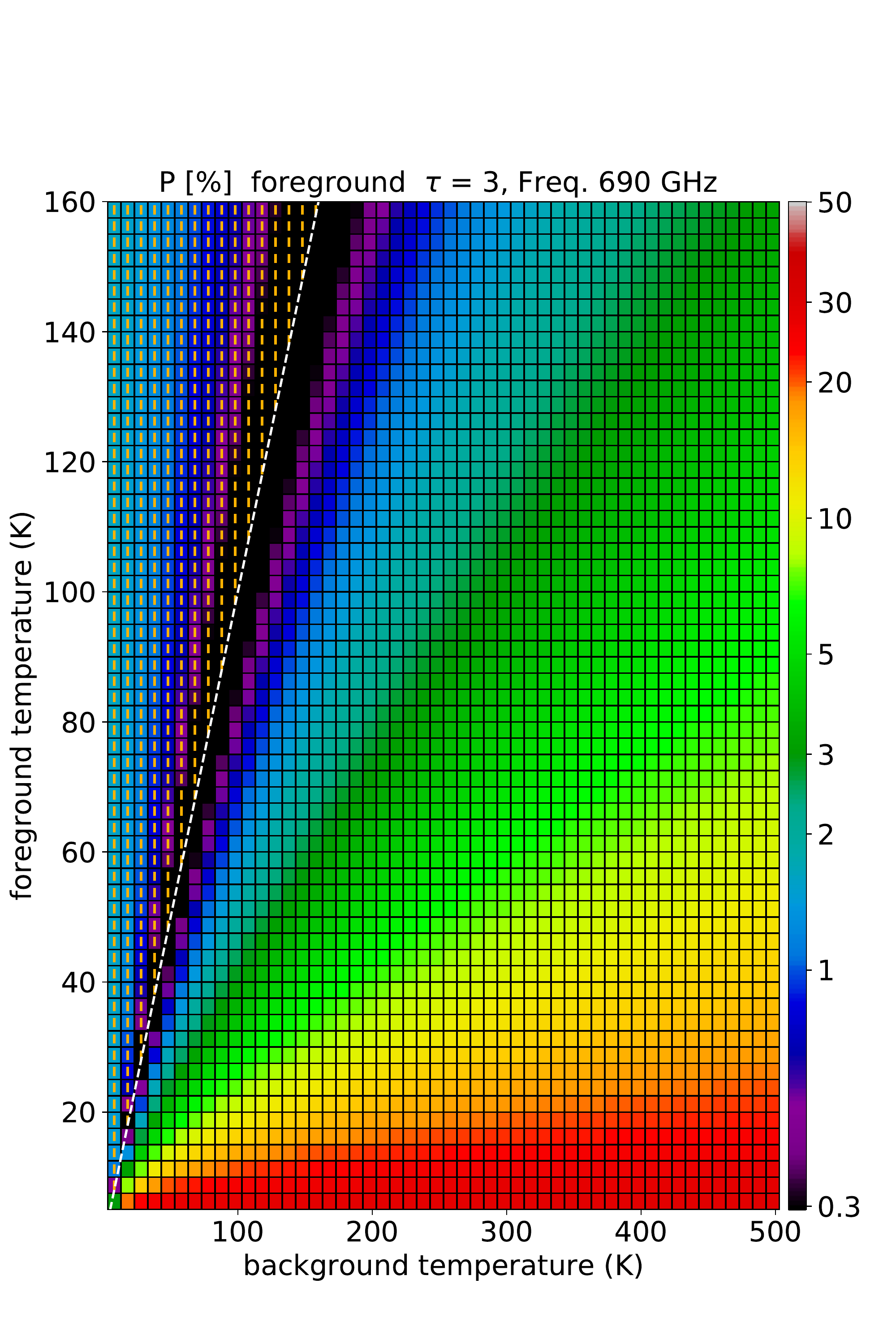}  \\
   \end{tabular}  

   \vspace{-0.3cm}
   \caption{
      \footnotesize{
      Similar to Figure \ref{fig:Permodel}, but for $\tau$=1 and 3 ($\alpha=10\%$).
      }
   }
   \label{fig:Permodel2}
\end{figure*}

\begin{figure*}
   \vspace{-1cm}
   \begin{tabular}{ p{5.6cm} p{5.6cm} p{5.6cm}  }
     \includegraphics[width=6.2cm]{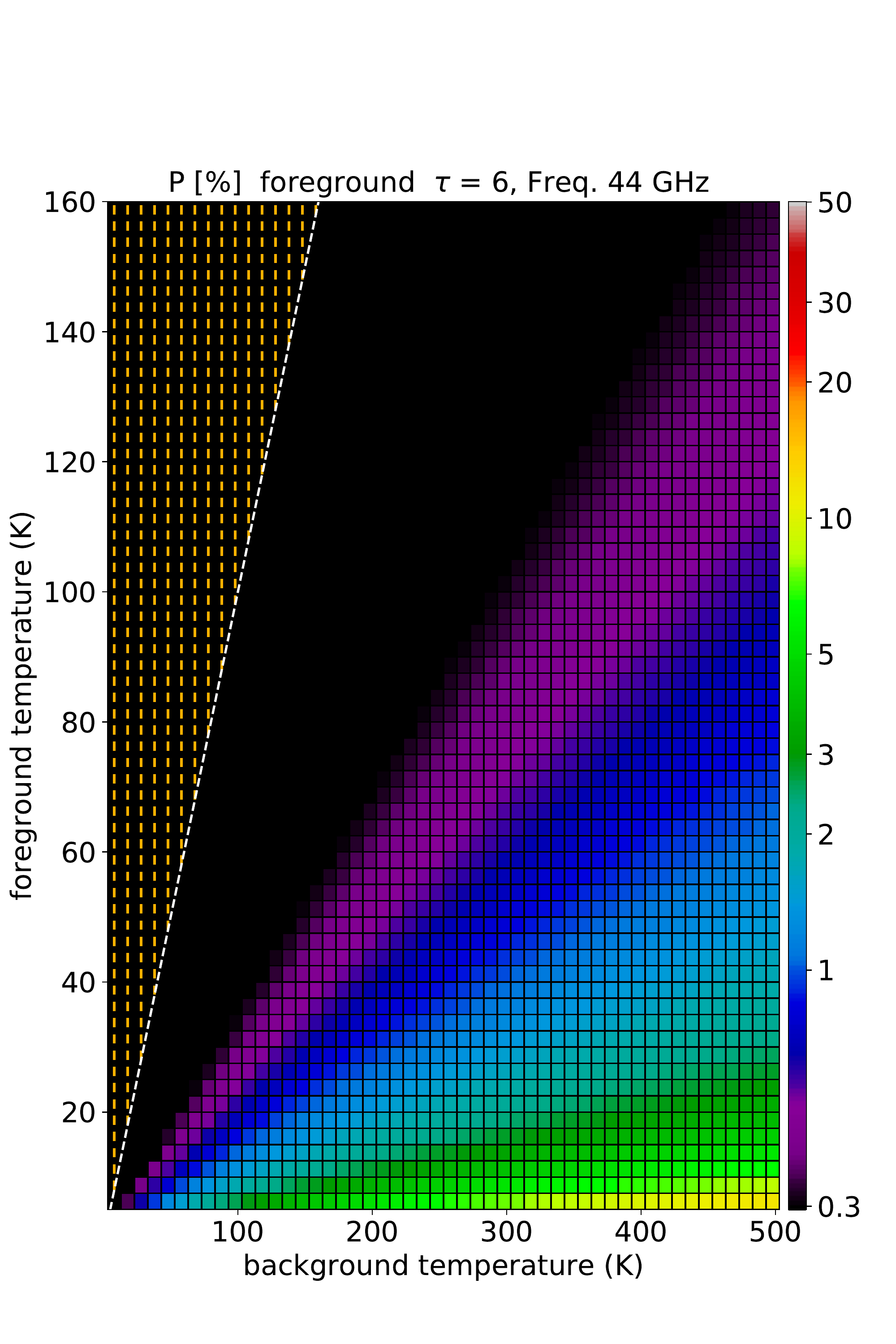} &
     \includegraphics[width=6.2cm]{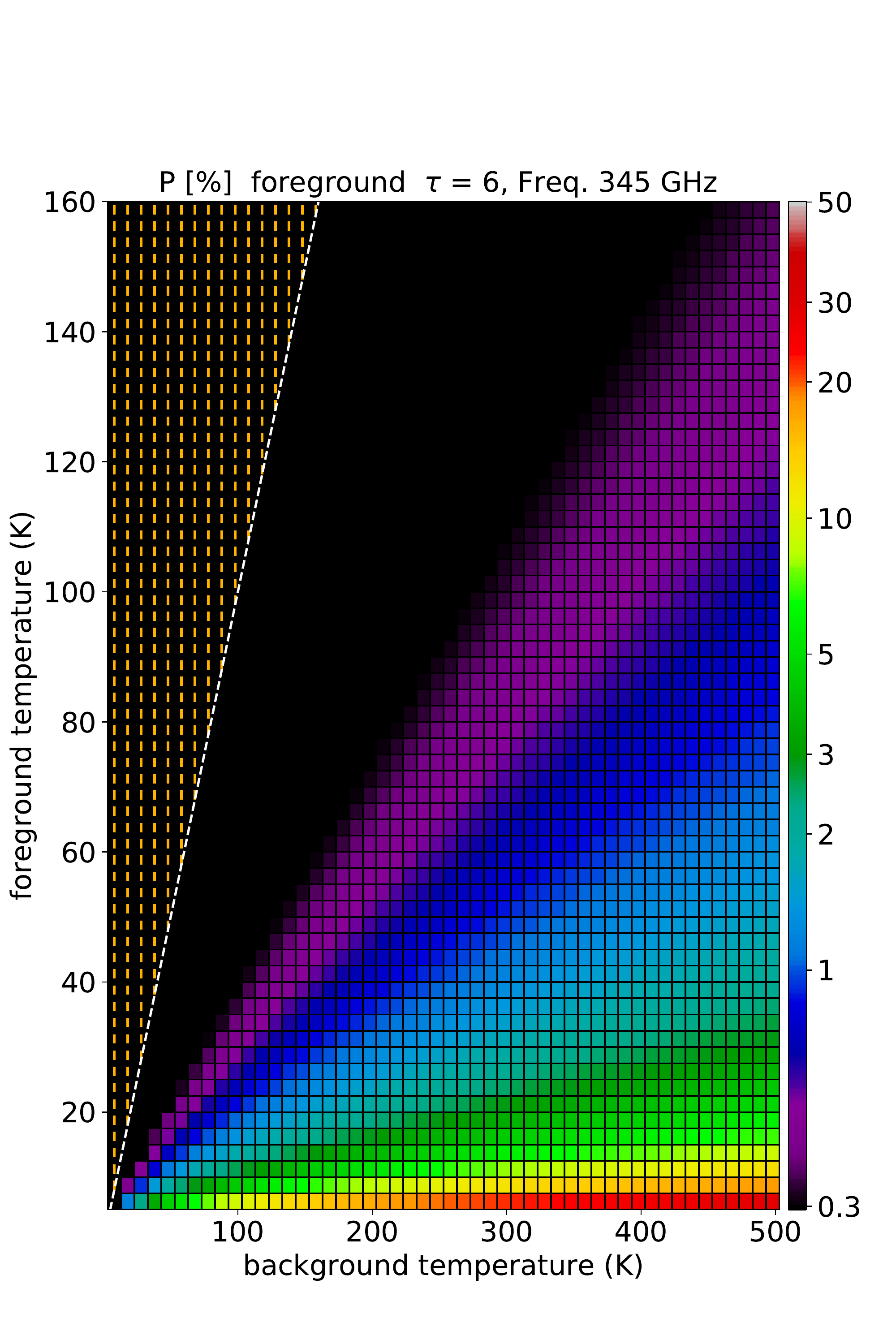} &
     \includegraphics[width=6.2cm]{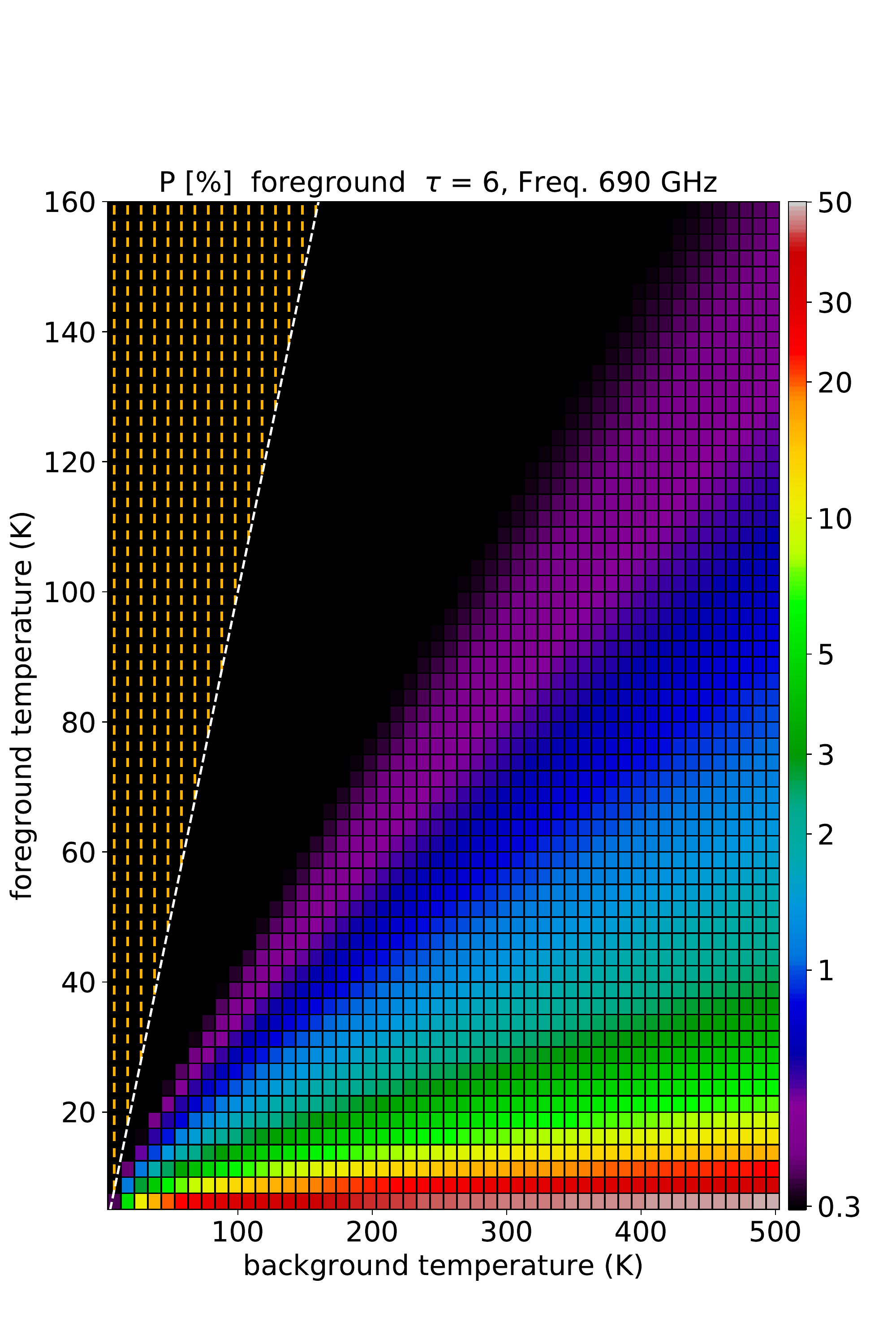}  \\
   \end{tabular}
   
   \vspace{-1.5cm}
   \begin{tabular}{ p{5.6cm} p{5.6cm} p{5.6cm}  }
     \includegraphics[width=6.2cm]{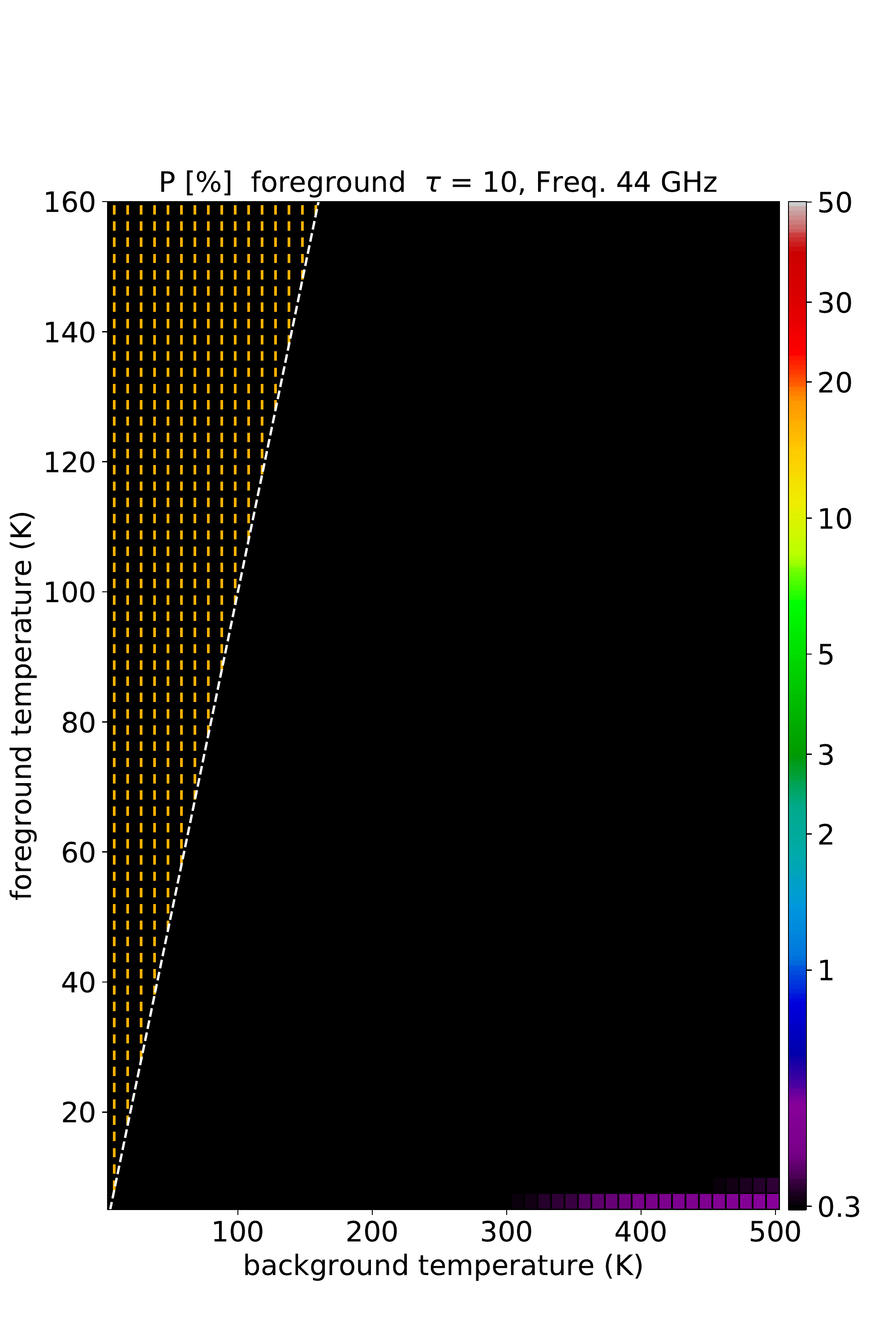} &
     \includegraphics[width=6.2cm]{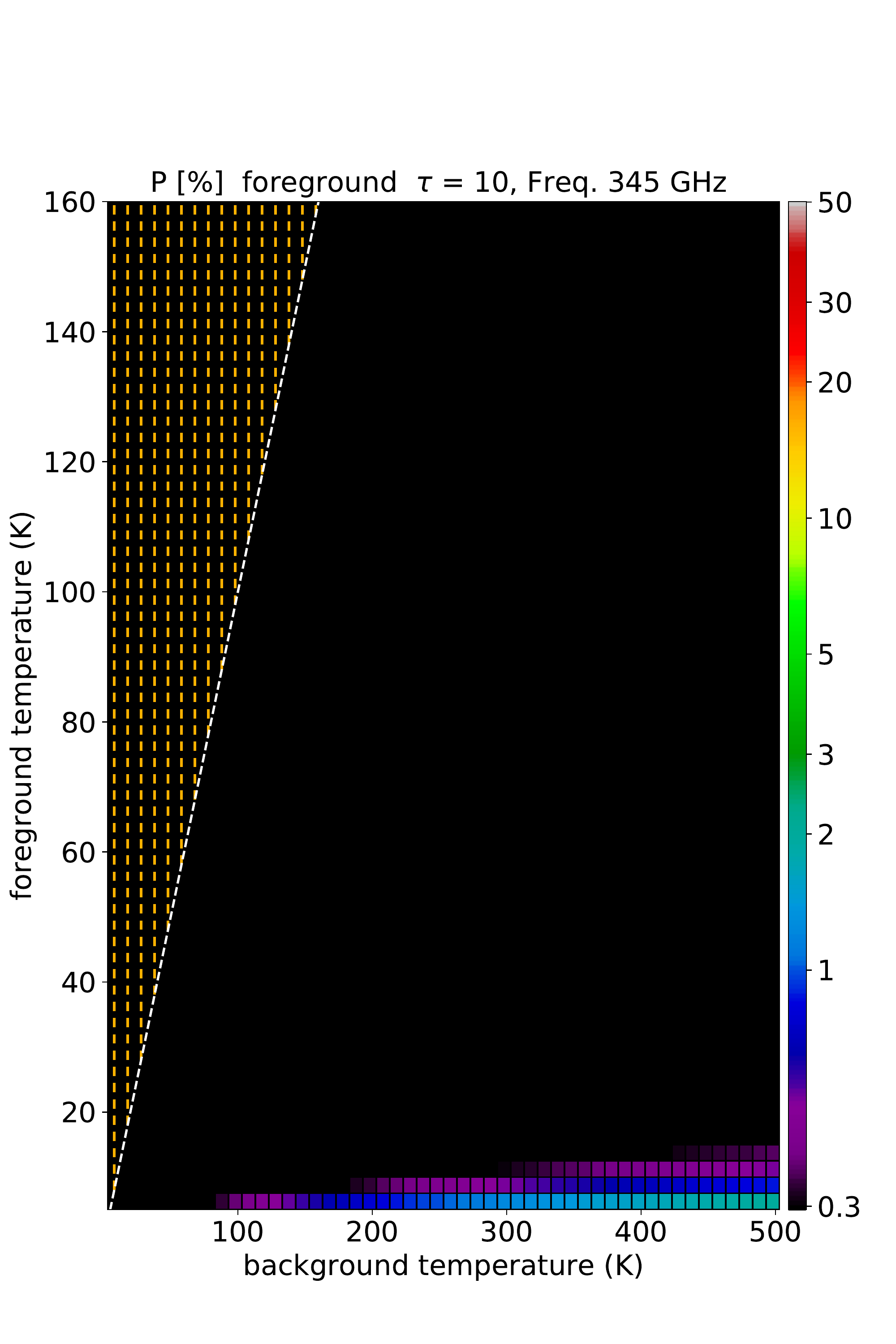} &
     \includegraphics[width=6.2cm]{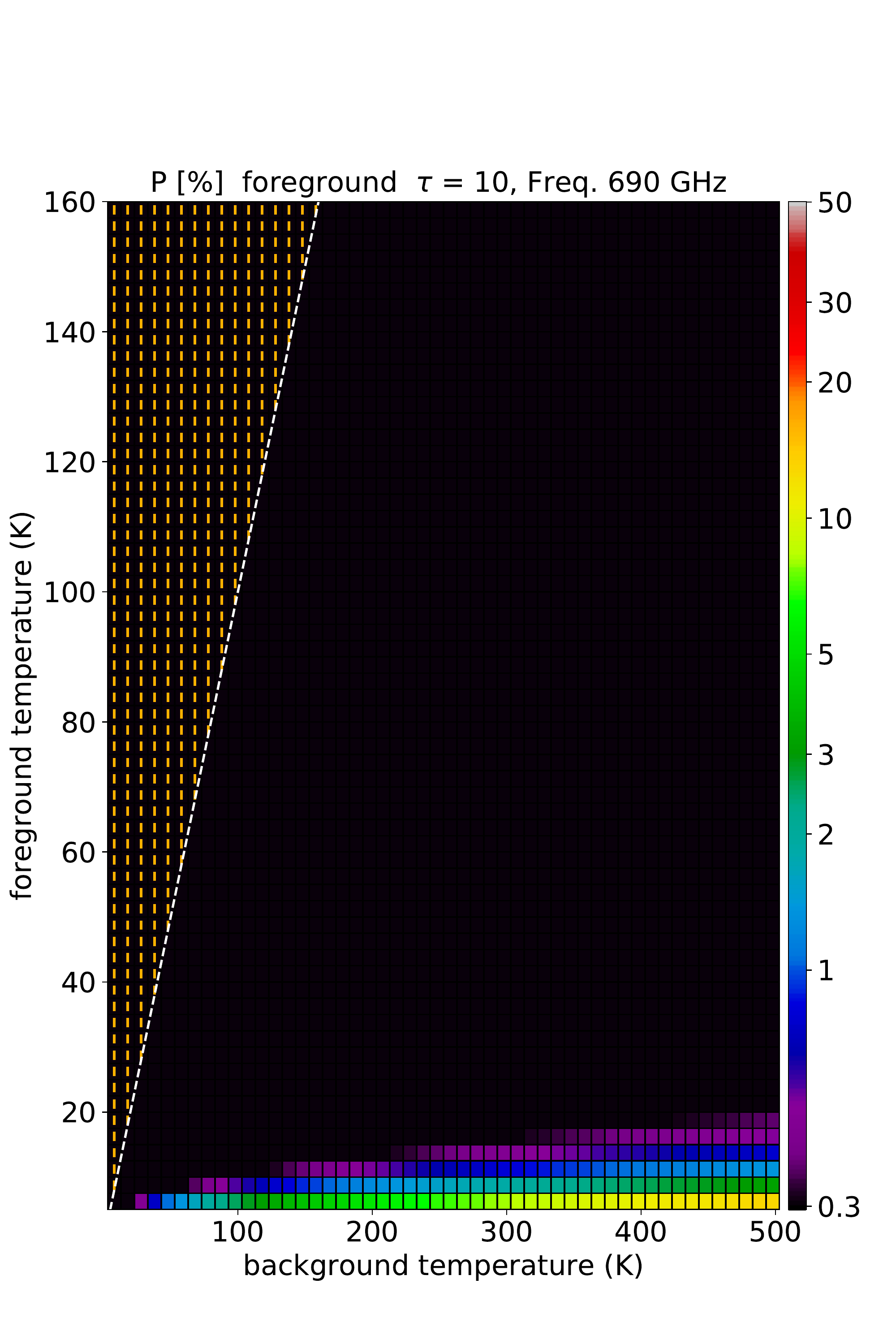}  \\
   \end{tabular}  
   \vspace{-0.3cm}
   \caption{
      \footnotesize{
      Similar to Figure \ref{fig:Permodel}, but for $\tau$=6 and 10 ($\alpha=10\%$).
      }
   }
   \label{fig:Permodel3}
\end{figure*}

\section{Observation and data reduction}\label{sec:observation}
We have performed JVLA Q band polarization observations toward IRAS\,16293-2422 in CnB array configurations on January 18 January 2015 (UTC 14:49:47.0-16:49:21.0; project code: 14B-053).
The observational setup was similar to that of the previous observations on the Class 0 object NGC\,1333\,IRAS4A, which were taken from the same JVLA project (Liu et al. 2016).
However, it is less easy to calibrate the observations of IRAS\,16293-2422 in general, and the polarization leakage term (D-term, hereafter) in particular, due to the low declination and thereby the low elevation (31.2$^{\circ}$ $\rightarrow$ 31.6$^{\circ}$ $\rightarrow$ 30.9$^{\circ}$ from beginning to transit to end) of the target source, the not very ideal local sidereal time (LST) for high-frequency operation during our observations, the very small covered parallactic angle ranges of all observed sources (e.g., $<$15$^{\circ}$ for the gain calibrator), and that the available low-polarization leakage calibrator for this particular target source was approximately 100 times fainter than the low-polarization leakage calibrator used for the NGC\,1333\,IRAS4A observations. 
Nevertheless, our observations and data calibration appear successful, which are described as follows.

The Q band receivers of JVLA use circular feeds, and therefore the Q and U parameters that define linear polarization are relatively unaffected by amplitude calibration errors (e.g., EVLA memo 113).
We took full RR, RL, LR, and LL correlator products. 
These observations had an overall duration of 120 minutes, with 46 minutes of integration on the target source.
After initial data flagging, 25 antennas were available. 
The projected baseline lengths covered by these observations are in the range of 77-4580 m ($\sim$11-670 $k\lambda$), which yielded a $\sim$0$\farcs$3 angular resolution ($\sim$44 au), and a maximum detectable angular scale (i.e., the recovered flux $\sim$1/e of the original flux) of $\sim$10$''$ ($\sim$1470 au; Wilner \& Welch 1994). 
We used the 3-bit sampler and configured the backend to have an 8 GHz bandwidth coverage by 64 consecutive spectral windows, which were centered on the sky frequency of 44 GHz. 
The pointing center for our target source is on R.A. = 16$^{\mbox{\tiny{h}}}$32$^{\mbox{\tiny{m}}}$22$^{\mbox{\tiny{s}}}$620 (J2000), decl. = -24$^{\circ}$28$'$32$''$50 (J2000). 
The complex gain calibrator J1625-2527 was observed for 54\,s every 222\,s to calibrate the atmospheric and instrumental gain phase and amplitude fluctuation. 
We observed the bright but weakly polarized ($<$1\%) quasar J2355+4950 for 594\,s for polarization leakage (i.e., D-term) calibrations. 
We integrated on the standard calibrator 3C286 for 465\,s for passband calibration and to reference the absolute flux scaling  and the polarization position angle. 
We further integrated on a likely very weakly polarized faint source J1407+2827 for 270\,s, to estimate the upper limit of residual polarization leakage after implementing the D-term solution derived from J2355+4950.
During our observations, we performed antenna reference pointing calibration when any calibrator source was observed for the first time.
We performed antenna reference pointing calibration on J1625-2527 at the beginning (UTC 15:19:58.0-15:23:04.0) and the middle (UTC 15:58:18.0-16:01:12.0) of the target source loop, which should have adequately suppressed the potentially relatively large pointing errors and the induced larger spurious polarization due to beam squint effect during the sunrise.

We manually followed the standard data calibration strategy using the Common Astronomy Software Applications (CASA; McMullin et al. 2007) package, release 5.1.1.
After implementing the antenna position corrections, weather information, gain-elevation curve, and opacity model, we bootstrapped delay fitting and passband calibrations, and then performed complex gain calibration, cross-hand delay fitting, polarization leakage calibration (using J2355+4950), and polarization position angle referencing. 
When solving D-term, spectral channels in each spectral window were pre-averaged to yield adequate signal-to-noise (S/N) ratios.
We applied the absolute flux reference to our complex gain solutions, and then applied all derived solution tables to the target source and J1407+2827.
We performed gain phase self-calibrations for our target source to remove the residual phase offsets, particularly, phase jumps.
Finally, we performed one iteration of gain amplitude self-calibration assuming that the antenna-based gain amplitude errors were constants during our observations.
The measured fluxes of J1407+2827, J1625-2527, and J2355+4950 at 43.97 GHz returned from the CASA task {\tt fluxscale} are 0.077$\pm$0.000097, 1.6$\pm$0.0035, and 0.23$\pm$0.00062 Jy, respectively.
We achieved an rms noise level of 77 $\mu$Jy\,beam$^{-1}$ in the Stokes Q, U and V images of J1407+2827, and did not find any detection.
The 3-$\sigma$ detection limits of the Stokes Q, U and V images of J1407+2827 constrain the residual polarization leakage to be $\lesssim$0.3\% after applying the D-term, cross-hand phase, and cross-hand delay solution. 
However, we caution that our measurements of polarization percentage from the target source can still be positively biased, due to the missing fluxes in the Stokes I intensity images. 


We generated images using the CASA task {\tt clean}. 
The image size is 3200 pixels in each dimension, and the pixel size is 0$\farcs$05. 
The achieved synthesized beams in the Briggs Robust = 0 and Robust = 2 weighted images are $\theta_{\mbox{\tiny{maj}}}$   $\times$ $\theta_{\mbox{\tiny{min}}}$ $=$ 0$\farcs$39 $\times$ 0$\farcs$24 (P.A. = 74$^{\circ}$) and $\theta_{\mbox{\tiny{maj}}}$   $\times$ $\theta_{\mbox{\tiny{min}}}$ $=$ 0$\farcs$49 $\times$ 0$\farcs$38 (P.A. = 66$^{\circ}$), respectively. 
The measured rms noise level is $\sim$35 $\mu$Jy\,beam$^{-1}$ in the Robust = 0 weighted Stokes I, Q, and U images, and is $\sim$30 $\mu$Jy\,beam$^{-1}$ in the Robust = 2 weighted images. 

We produced polarization intensity images, polarization percentage images, and polarization position angle images using the Miriad task {\tt impol} (Sault et al. 1995), and set the signal-to-noise (S/N) ratio cut off to be 3.0 to suppress the positive bias in the polarization intensity images (see discussion in Vaillancourt 2006).

\section{Results}\label{sec:results}
Figure \ref{fig:jvla} shows the Robust = 2 and Robust = 0 weighted total intensity (Stokes I), polarization intensity (PI), and polarization position angle (P.A.) images, which were produced from our JVLA 40-48 GHz continuum observations.
In Figure \ref{fig:jvla}, the line segments representing our polarization position angle (E-field) measurements are hexagonally packed and are separated by 0.4 times the median synthesized beam size ($\theta_{\mbox{\tiny{maj}}}$ + $\theta_{\mbox{\tiny{min}}}$)/2.
The Stokes I images resolve the two binary components, namely IRAS\,16293-2422\,A and B.
Our 40-48 GHz observations detect dust polarization at position angles and percentages very similar to those of the SMA observations at 341.5 GHz reported in Rao et al. (2009; 2014).
Our quantitative measurements are listed in Appendix \ref{appendix:poltable}.
We caution that for regions with low Stokes I intensities, the measurements of polarization percentage may be biased by missing Stokes\,I flux (see Section \ref{sub:interpretation} for more discussion).
The Robust = 0 images only detect polarization intensity from the inner region of IRAS\,16293-2422\,B, and therefore the presentation is limited to a small area around it.
The lower angular resolution JVLA images generated by tapering are provided in Appendix \ref{appendix:taper}, which however should not be used in any quantitative studies due to low S/N ratios and poor image fidelity.

From the Robust = 0 weighted image, the observed peak flux density of Stokes I continuum emission is 21 mJy\,beam$^{-1}$ ($\sim$141 K), detected from the IRAS\,16293-2422\,B component. 
The previous VLA observations towards IRAS\,16293-2422\,B (e.g., Chandler et al. 2005; Rodr{\'{\i}}guez et al. 2005) have shown that dust emission is dominant even at the frequency of $\sim$8.5 GHz.
In this case, it is not trivial to observationally constrain the fluxes of free-free continuum emission at our present observing frequency $\sim$44 GHz.
Nevertheless, based on the previously reported fluxes at $\sim$4.8 GHz (Wooten 1989; Chandler et al. 2005; Rodr{\'{\i}}guez et al. 2005), we expect that free-free emission contributed to $\ll$1\% of the observed 44 GHz intensity around IRAS\,16293-2422\,B, which can be safely neglected for our following analysis.

We note that the peak brightness temperature of our target source at 44 GHz (6.9 mm) is comparable with that report from the ALMA b$_{\mbox{\tiny{maj}}}$   $\times$ b$_{\mbox{\tiny{min}}}$ $=$ 0$\farcs$32 $\times$ 0$\farcs$18, observations at 0.45 mm ($\sim$160 K; Loinard et al. 2013; see also Loinard et al. 2007).
Here we quote an empirical and nominal $\sim$20\% absolute flux calibration error for ALMA observations at 0.45 mm wavelength.
This comparison suggests that dust continuum emission from the inner $\sim$50 au diameter of IRAS\,16293-2422\,B is likely optically very thick at $<$6.9 mm wavelengths.
Our tentative interpretation for the polarization mechanisms at both frequencies will be discussed in the following section.
We refer to Flett et al. (1991), Tamura et al. (1993), and Akeson \& Carlstrom(1997) for earlier (sub)millimeter observations on our target source, and note that we are not yet able to consistently interpret all of them with the observations of ours and of Rao et al. (2009; 2014).
The discrepancy may be related to missing short-spacing (see Appendix \ref{appendix:line} for more discussion) or other calibration effects that we cannot trace back.

\begin{figure*}
   \vspace{-0.3cm}
   \begin{tabular}{ p{4.2cm} p{4.2cm} p{4.2cm} p{4.2cm} }
     \includegraphics[width=4.2cm]{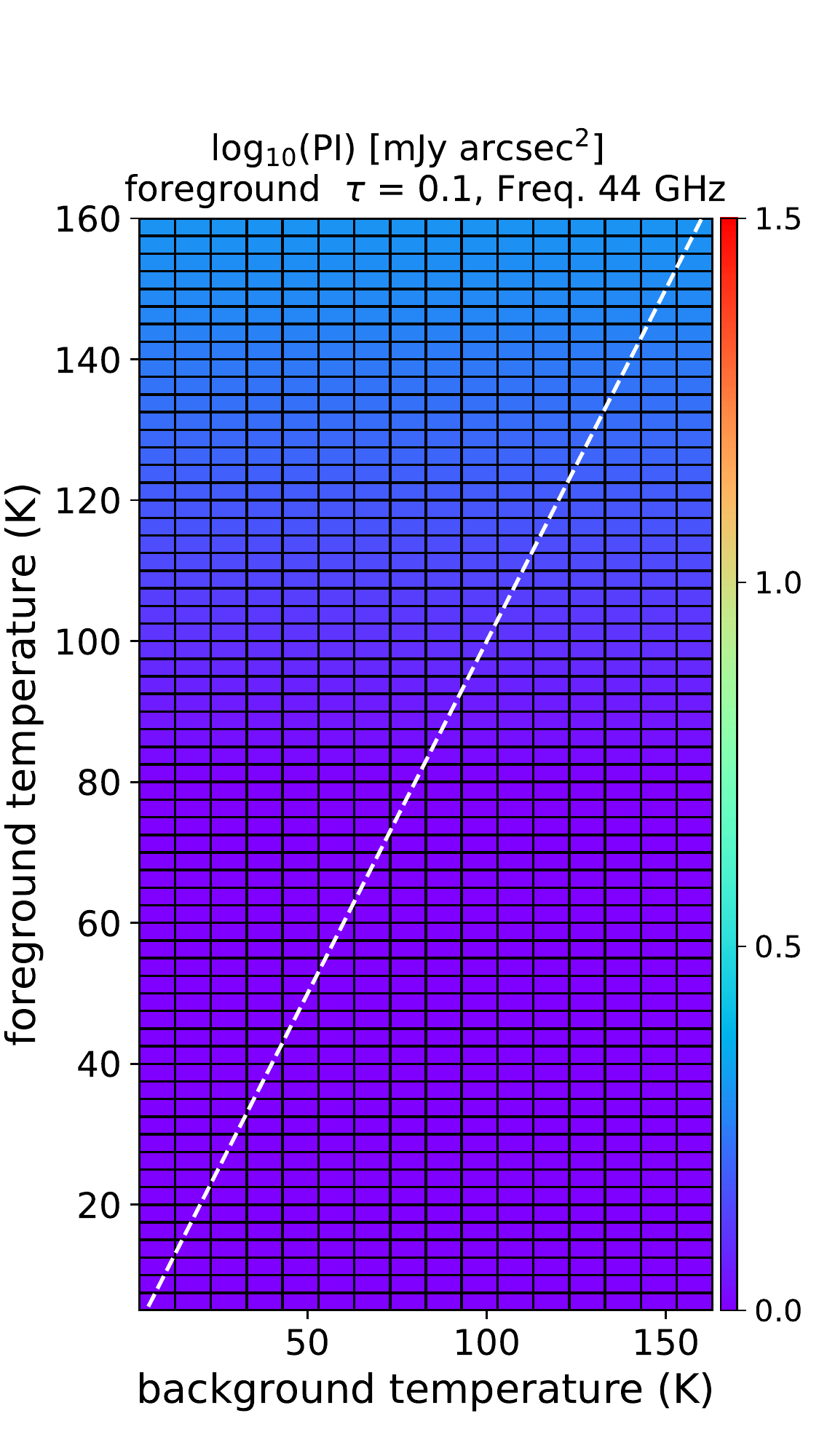} &
     \includegraphics[width=4.2cm]{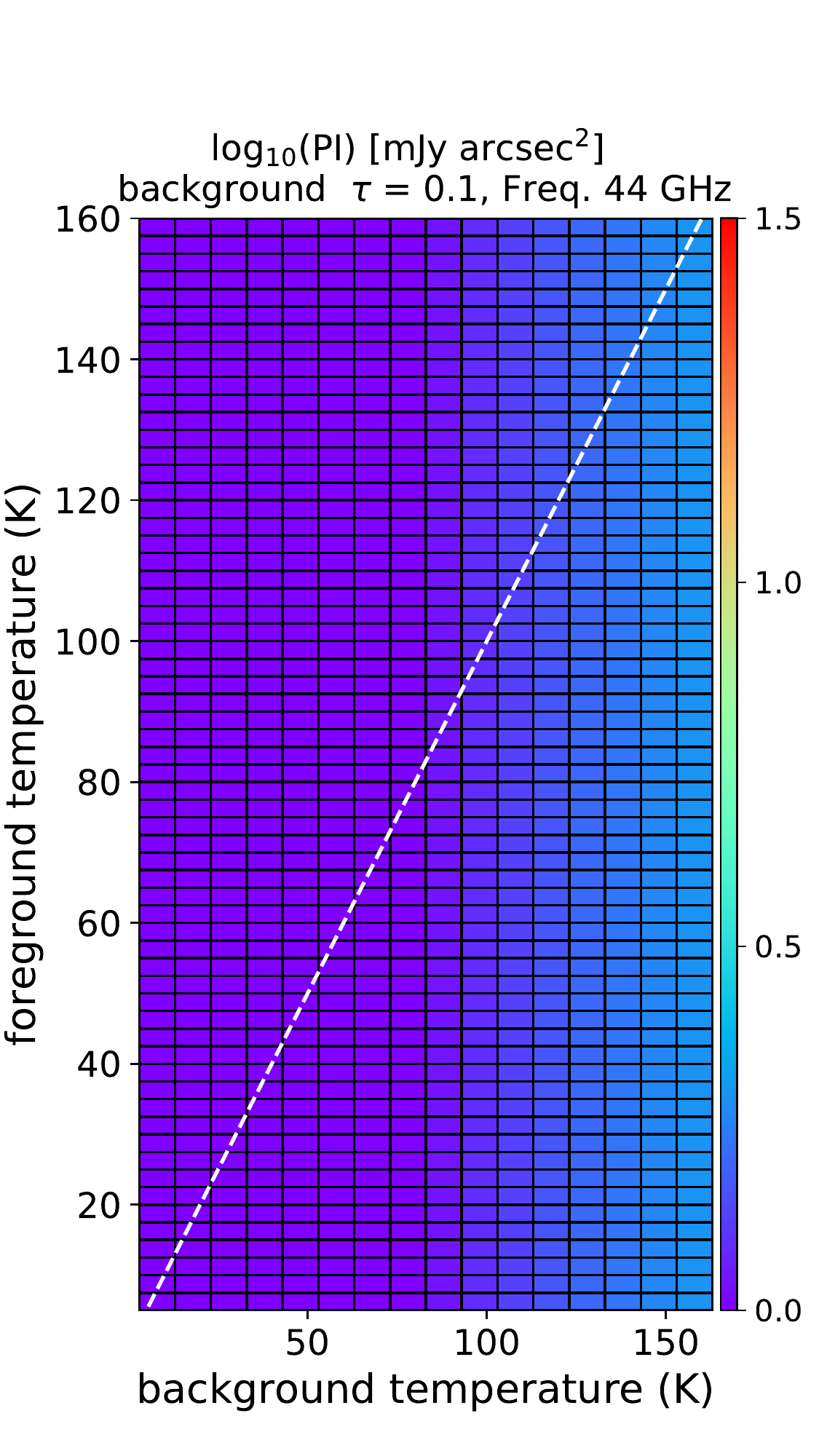} &
     \includegraphics[width=4.2cm]{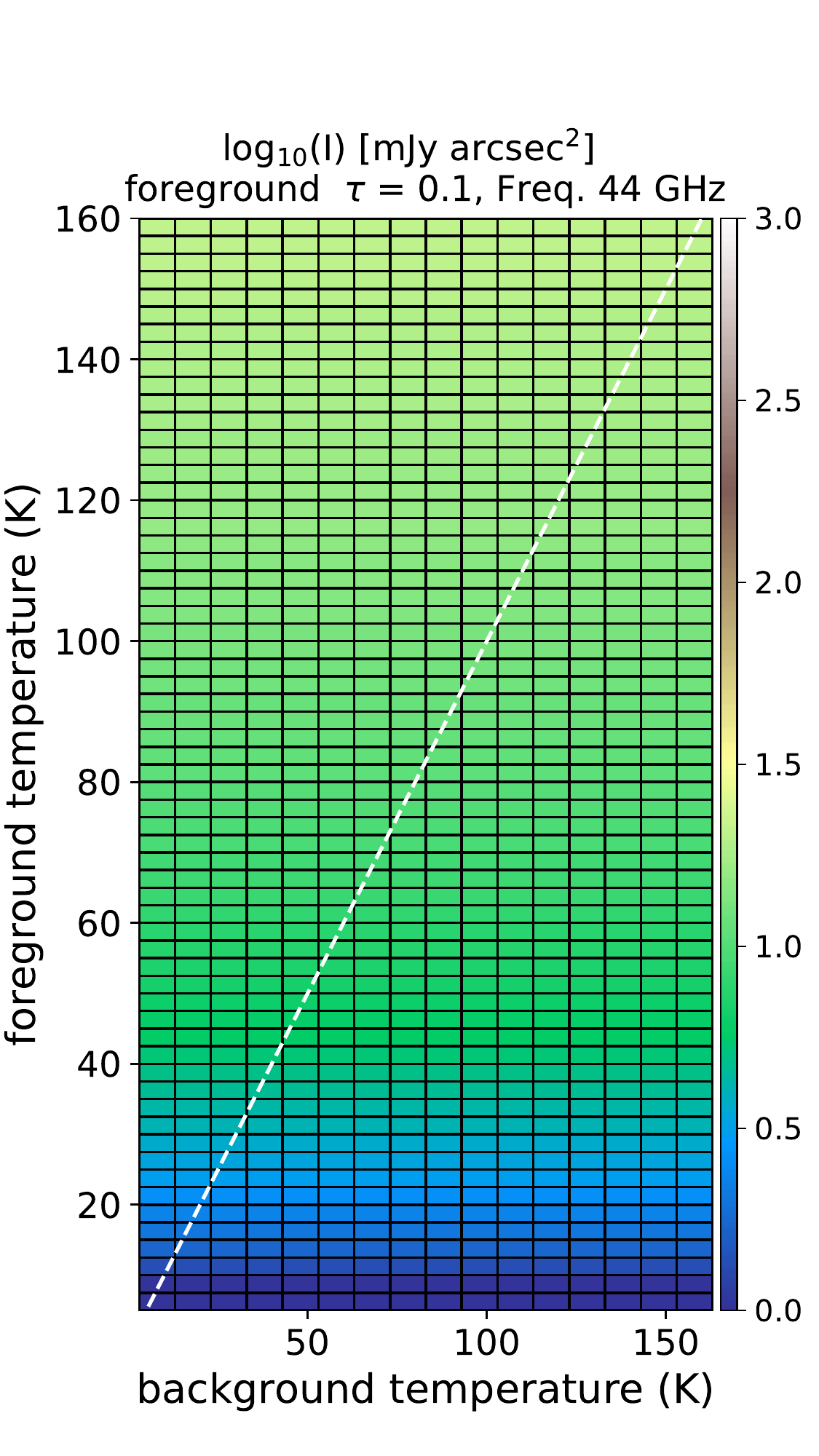} &
     \includegraphics[width=4.2cm]{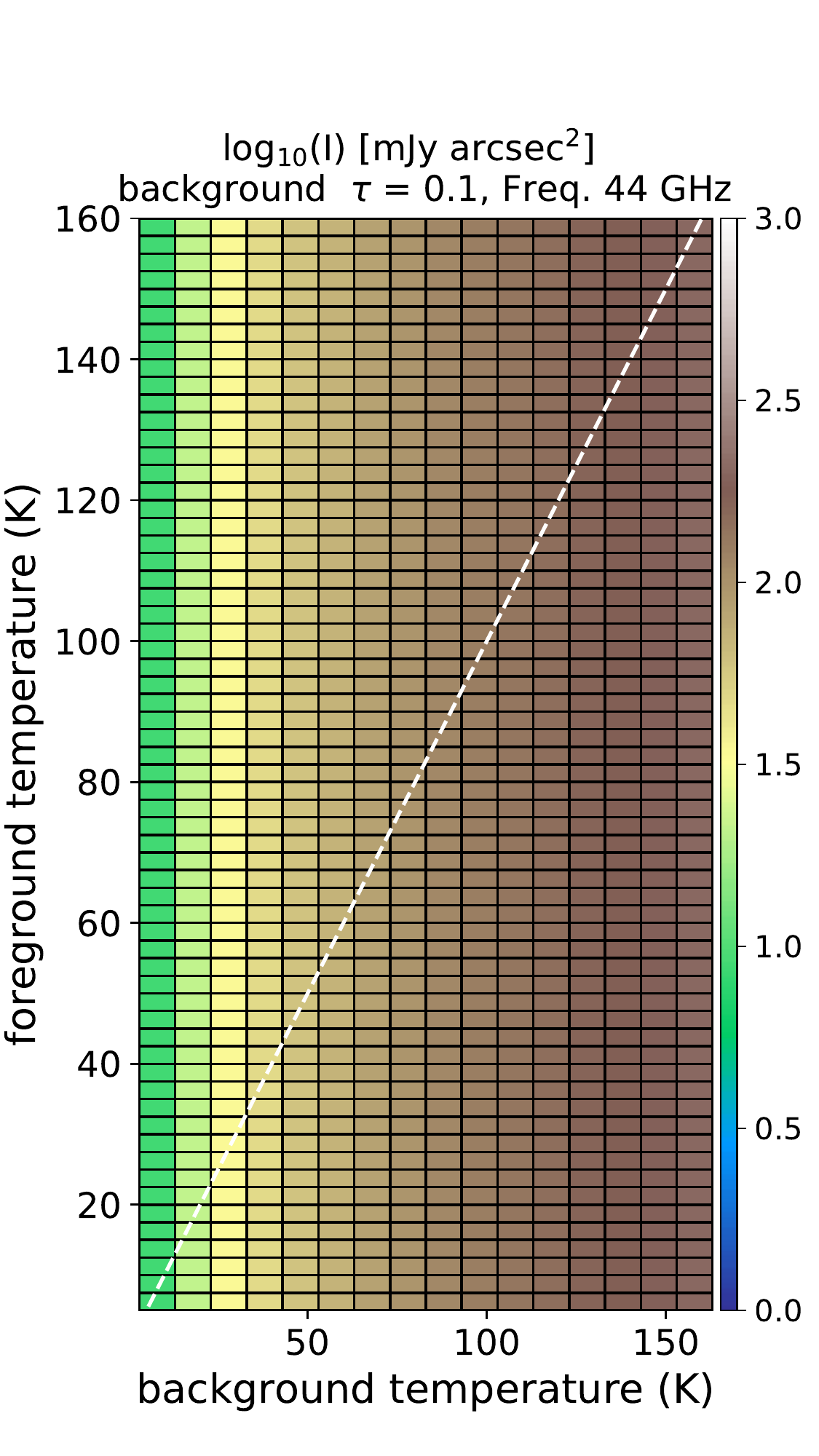}  \\
   \end{tabular}
   
   \vspace{-0.82cm}
   \begin{tabular}{ p{4.2cm} p{4.2cm} p{4.2cm} p{4.2cm} }
     \includegraphics[width=4.2cm]{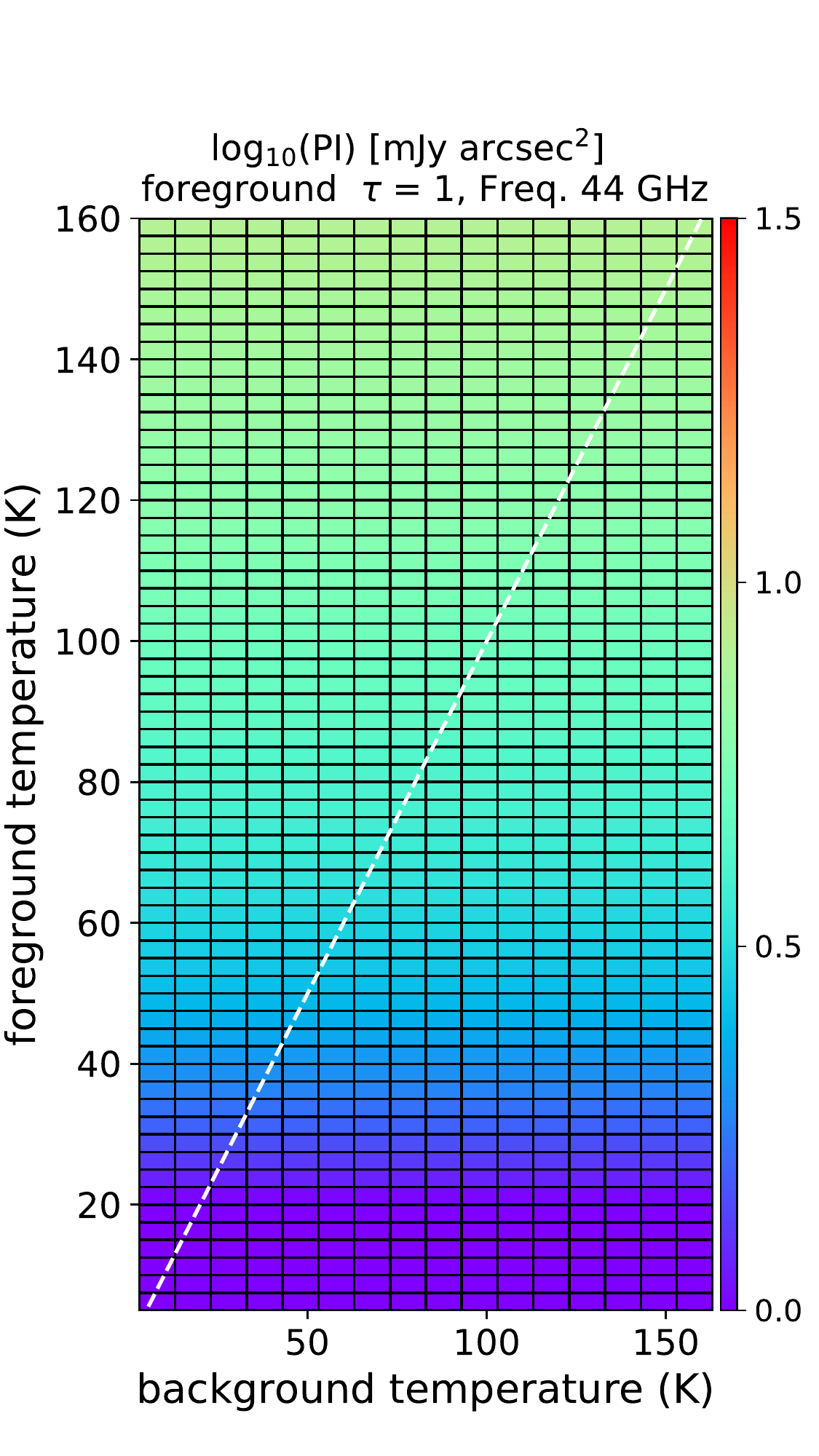} &
     \includegraphics[width=4.2cm]{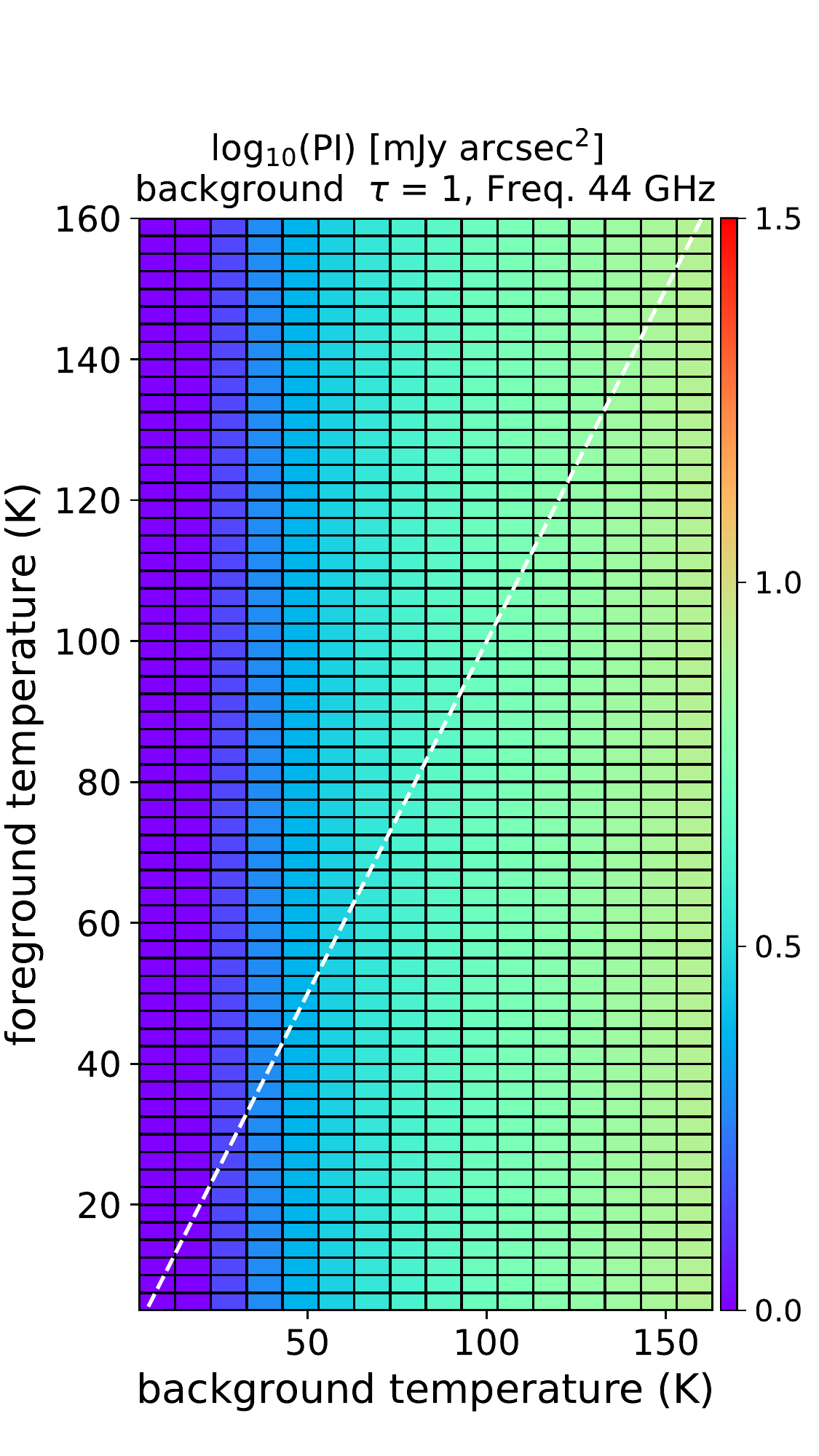} &
     \includegraphics[width=4.2cm]{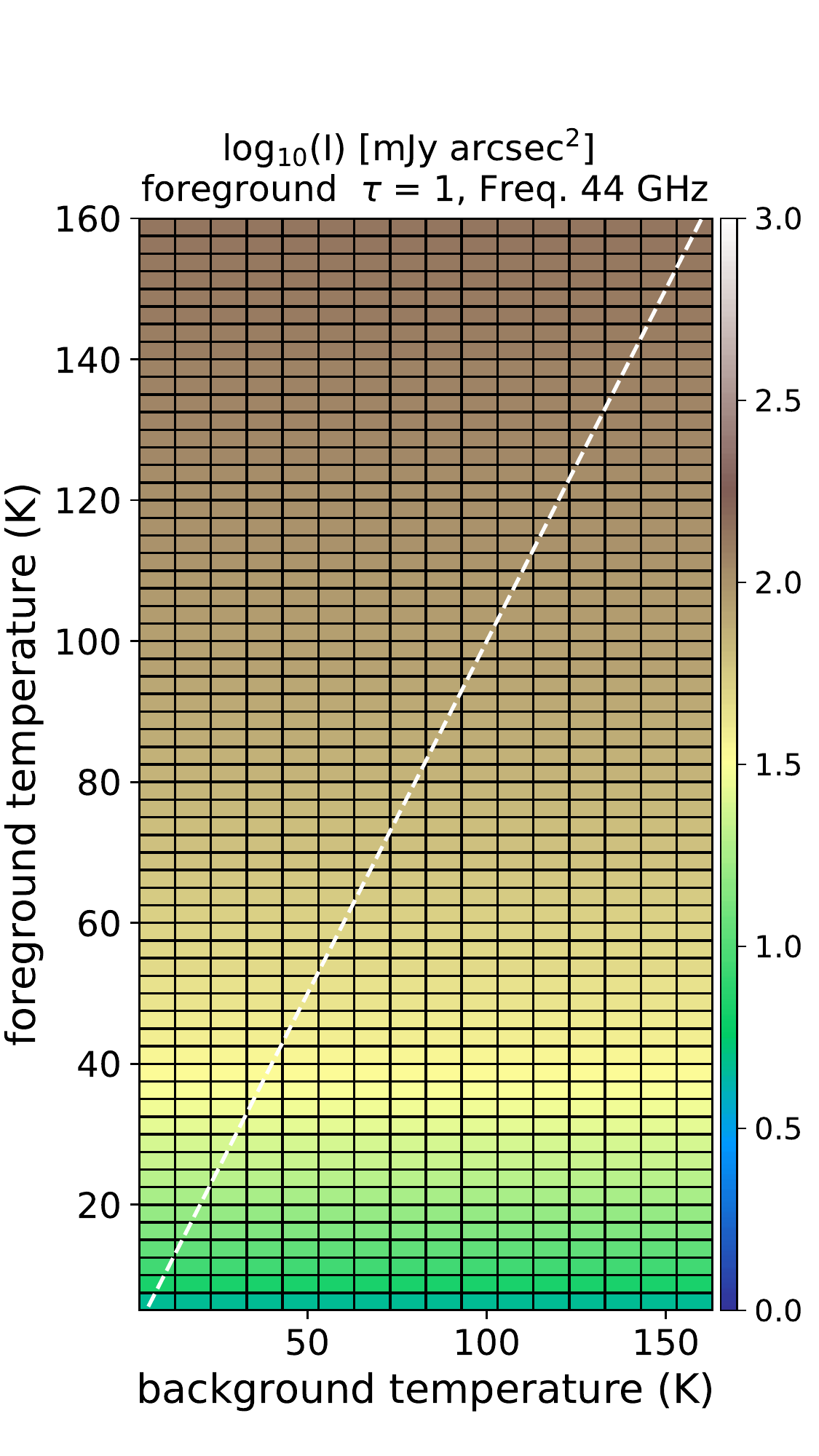} &
     \includegraphics[width=4.2cm]{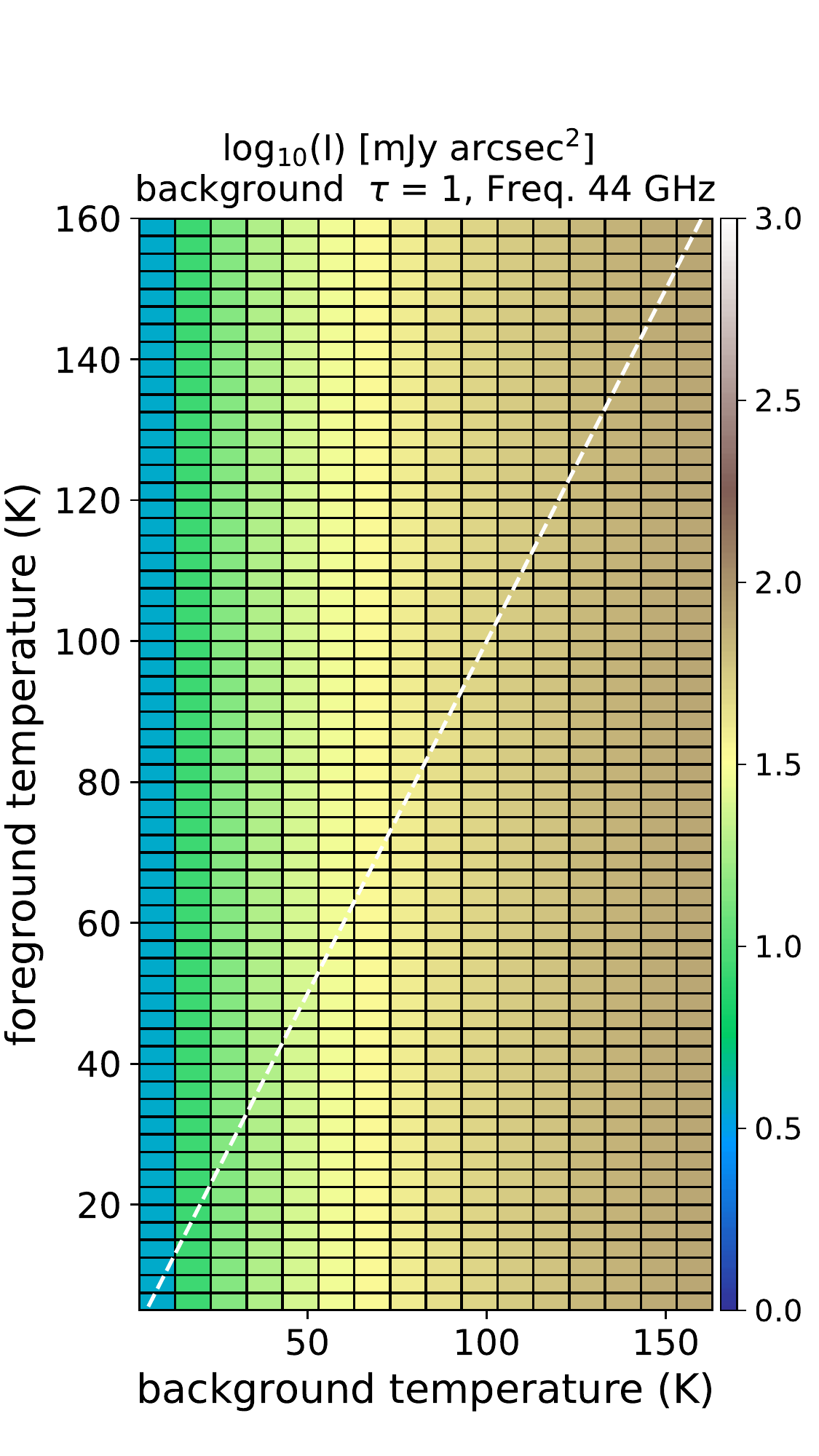}  \\
   \end{tabular}
   
   \vspace{-0.82cm}
   \begin{tabular}{ p{4.2cm} p{4.2cm} p{4.2cm} p{4.2cm} }
     \includegraphics[width=4.2cm]{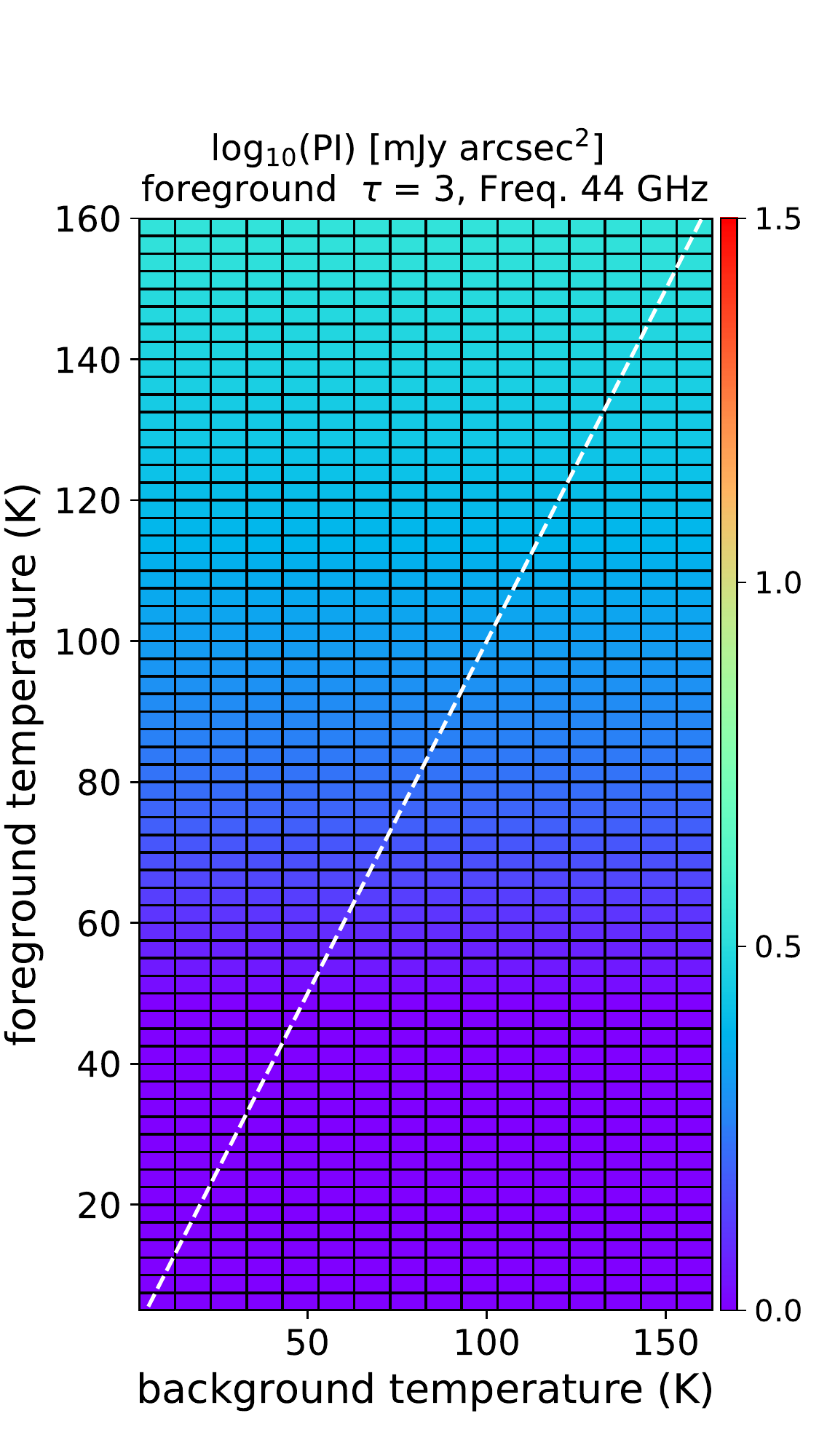} &
     \includegraphics[width=4.2cm]{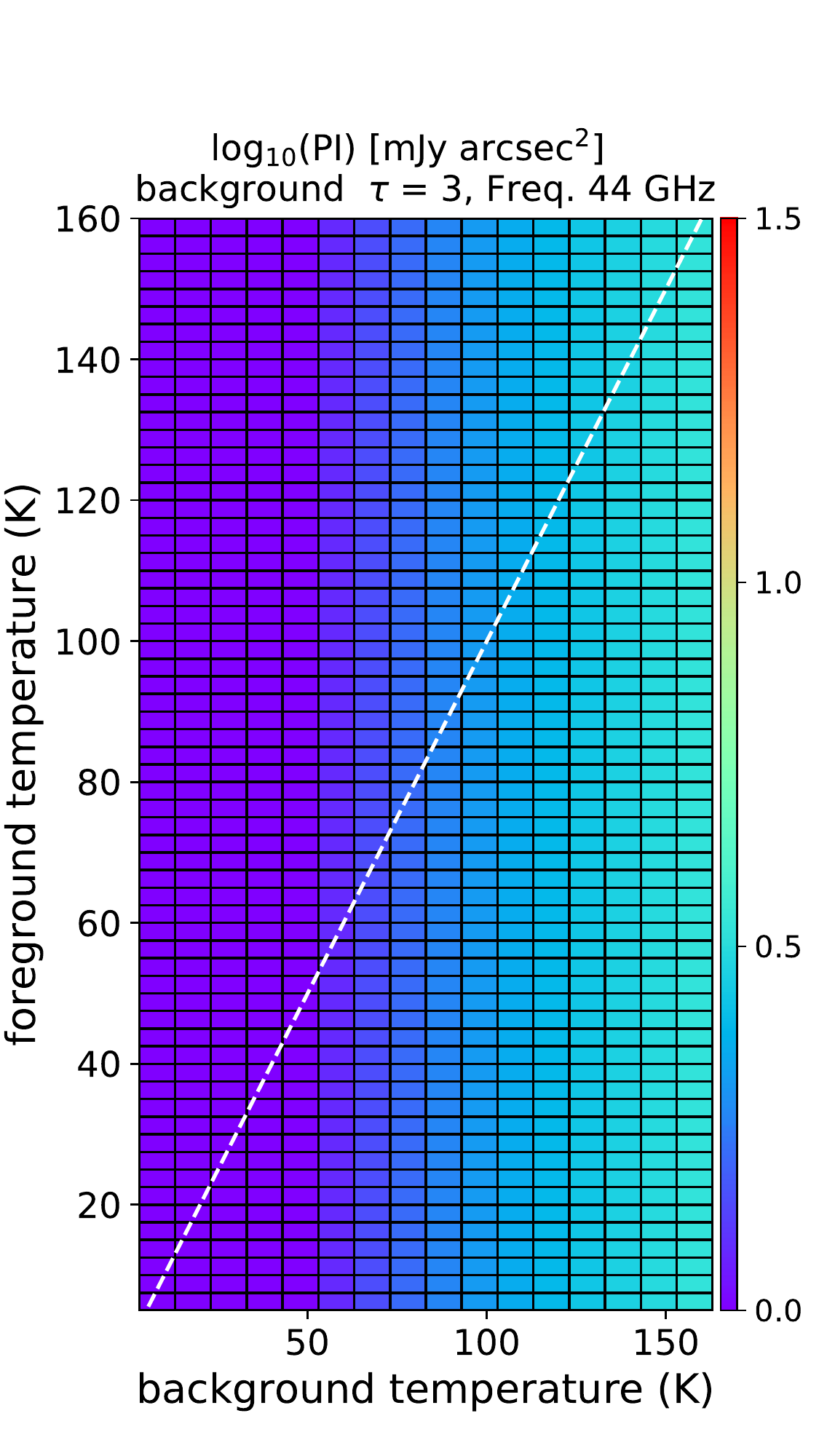} &
     \includegraphics[width=4.2cm]{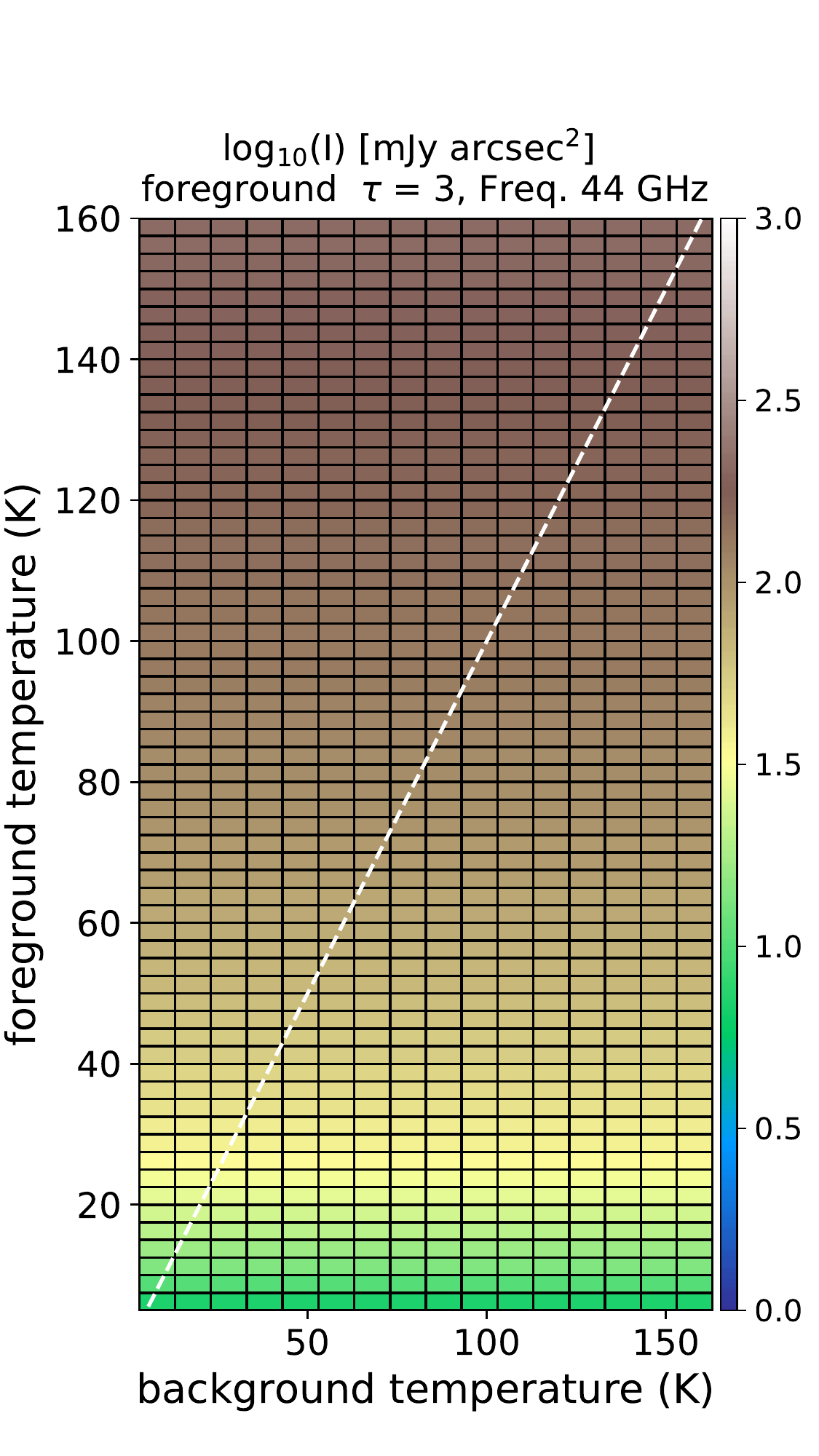} &
     \includegraphics[width=4.2cm]{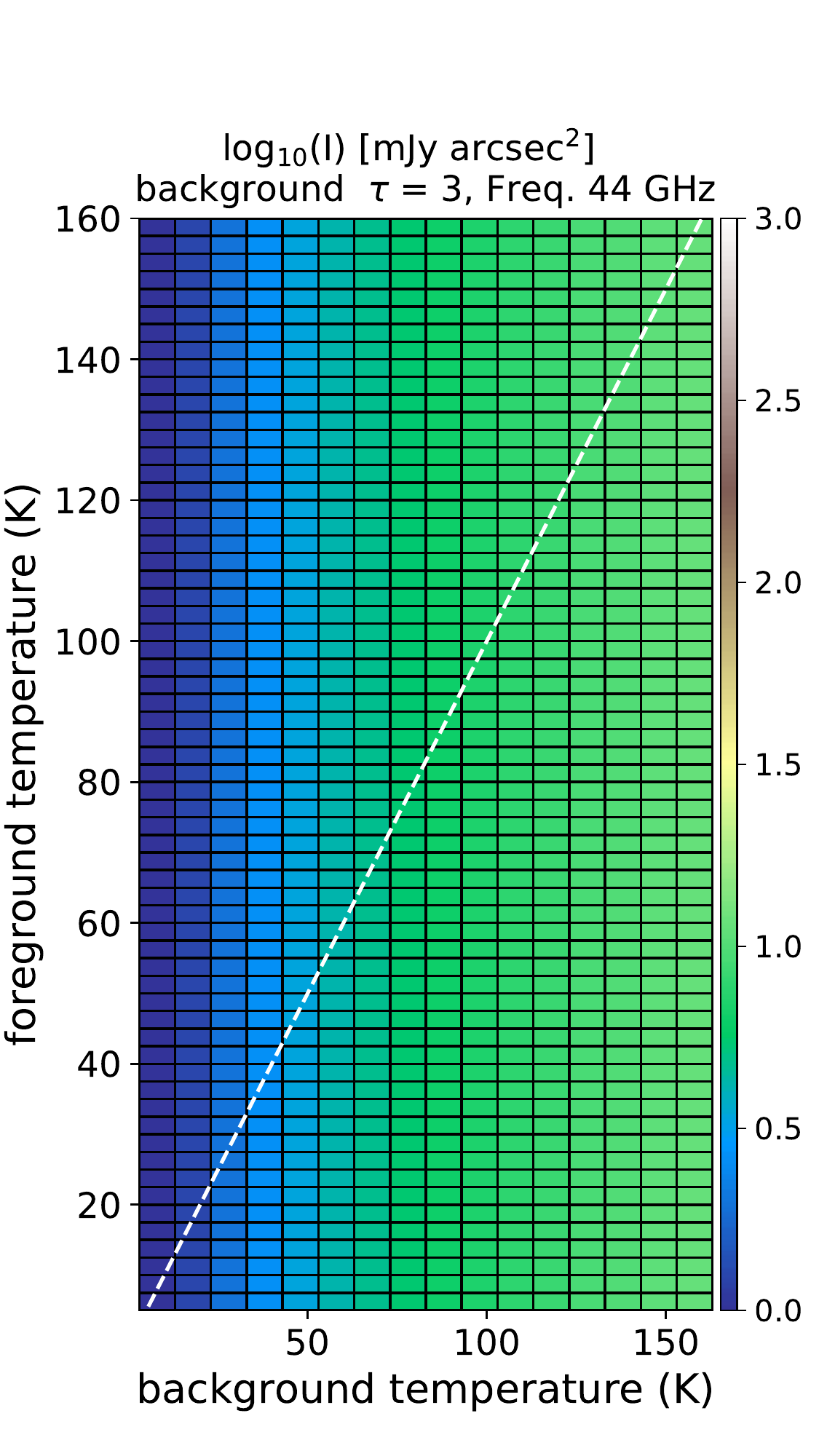}  \\
   \end{tabular}
   \vspace{-0.35cm}
   \caption{
      \footnotesize{
      Derived polarization (PI) and total intensities (I) from the two-component model introduced in Section \ref{sub:mechanism} (Equation \ref{eqn:multicomponent}), with $\alpha=10\%$ (Equation \ref{eqn:alpha}), and observing frequency $\nu=$44 GHz. Color bars are in mJy\,beam$^{-1}$ units. Horizontal and vertical axes are dust temperatures of the background and foreground components $T_{fg}$ and $T_{bg}$. White dashed lines show $T_{fg}=T_{bg}$. From top to bottom rows show the cases of $\tau=$0.1, 1 and 3, respectively.
      }
   }
   \label{fig:PImodel}
\end{figure*}

\section{Discussion}\label{sec:discussion}
Around star-forming cores or circumstellar disks, the probable and non-mutually exclusive mechanisms to linearly polarize dust continuum emission include (1) aligning elongated dust grains with magnetic field (e.g., Hildebrand et al. 2000), (2) radiative torque alignment of asymmetric dust grains by anisotropic emission field (e.g., Tazaki et al. 2017), (3) mechanical alignment (e.g., Lazarian et al. 1996), and (4) scattering of anisotropic radiation field (e.g., Kataoka et al. 2015).

For the particular Class 0 YSO IRAS\,16293-2422, the previous chemical modeling based on the molecular abundance ratios reported by Chandler et al. (2005) suggested that the maximum grain size is in the range of 10-100 $\mu$m (Harada et al. 2017).
In addition, the previous studies of (sub-)millimeter Stokes I spectral indices from a group of Class\,0 YSOs found no evidence about grain growth (Li et al. 2017).
Therefore, we argue that on the spatial scale and wavelength range probed by our JVLA observations and the previous SMA polarization observations, dust scattering opacity may be too low (see Kataoka et al. 2015 for more estimates), such that scattering may not be a significant polarization mechanism (c.f. Alves et al 2018).
Scattering may become more prominent at shorter wavelength and in the inner $\ll$50 au region of IRAS\,16293-2422\,B, where the maximum grain size is not yet well constrained by any previous observations due to the very high optical depth of dust (Rodr{\'{\i}}guez et al. 2005).

We are not yet able to discern case (1), (2), or (3).
Individuals of these mechanisms may be prominent on different spatial (and/or dynamic time) scales or in different physical environments.
In Section \ref{sub:mechanism} we elaborate on the resulting polarization in optically thick and optically thin regime without distinguishing the grain alignment mechanisms.
Tentatively, we are favoring case (1) since in star-forming regions, the observed polarization position angles are often well connected over a wide range of spatial scales (for a review see Li et al. 2014, and references therein).
Having this in mind, and based on the quantitative estimates in Section \ref{sub:mechanism}, our interpretation for the specific case of IRAS\,16293-2422\,B is briefly described in Section \ref{sub:interpretation}.
We caution that in any case our present interpretation is likely oversimplified.
Understanding radiative alignment and dust scattering requires full three-dimensional radiative transfer modeling in the high and low optical depth regimes, which is well beyond our expertise.
In addition, in the optically thin regime, the polarized dust emission from aligned dust grain   may be subject to non-trivial line-of-sight canceling when the field configurations are very complicated.
We refer to Kataoka et al. (2012) and Yang et al. (2017), and references therein, for the more advanced modeling frameworks.

\subsection{Polarization mechanism by aligned dust grains}\label{sub:mechanism}
Limited by atmospheric transmission, sensitivity, and the calibration difficulties of the present generation instruments (e.g., James Clerk Maxwell Telescope, Combined Array for Research in Millimeter-wave Astronomy, SMA, JVLA), most of the existing dust polarization measurements towards circumstellar disks and envelopes targeted on warm and bright sources at millimeter bands, which are well in the Rayleigh-Jeans limit.
We expect this to remain true in the upcoming decade, with only some exceptional cases which perform deep integration towards low-temperature starless cores with extremely accurate polarization calibration, and some observations using the High-resolution Airborne Wideband Camera-plus (HAWC+) of the Stratospheric Observatory for Infrared Astronomy (SOFIA) towards very nearby sources (e.g., Novak 2016).
Given that in the Rayleigh-Jeans limit, flux of dust emission is not sensitive to temperature, modeling a very detailed three-dimensional temperature distribution may not be necessary.
Therefore, in the following discussion, we consider a highly simplified yet comprehensive scenario which includes only two dust components in a line-of-sight.
In addition, we assume that individual of these two dust components are isothermal.

For one line-of-sight, the dust fluxes in a locally and arbitrarily defined orientation (E), and the orientation (B) which is orthogonal to E may be approximated by Equation \ref{eqn:multicomponent}.

\begin{equation}\label{eqn:multicomponent}
   \begin{split}
      2F^{E}_{\nu} =  \Omega_{bg}(1-e^{-\tau^{E-bg}_{\nu}})B_{\nu}(T_{bg})e^{-\tau^{E-fg}_{\nu}} + 
           \Omega_{fg}(1-e^{-\tau^{E-fg}_\nu})B_{\nu}(T_{fg}) \\
      2F^{B}_{\nu} =  \Omega_{bg}(1-e^{-\tau^{B-bg}_{\nu}})B_{\nu}(T_{bg})e^{-\tau^{B-fg}_{\nu}} + 
           \Omega_{fg}(1-e^{-\tau^{B-fg}_\nu})B_{\nu}(T_{fg})
   \end{split}
\end{equation}
where $\Omega_{fg, bg}$ are the solid angles of the foreground and background components, $B_{\nu}(T)$ is the Planck function for a certain temperature ($T$) of the source, and $T_{fg, bg}$ are the dust temperatures of these components.
Finally, $\tau^{E-bg}_{\nu}$, $\tau^{B-bg}_{\nu}$, $\tau^{E-fg}_{\nu}$, $\tau^{B-fg}_{\nu}$ are the dust optical depths of the background and foreground component in the orthogonal orientations E and B, at the observing frequency $\nu$. 
For high angular resolution observations towards Class 0/I YSOs specifically, the background and foreground components may correspond to the central part of the accretion disk (or disk-like structure), and the outer part of the disk(-like structure) or the inner envelope, respectively.
Equation \ref{eqn:multicomponent} is essentially a polarized generalization of Equation 6 in Li et al. (2017), but drops the free-free emission term for simplicity.
We refer to Li et al. (2017) for the implications about the Stokes\,I spectral slopes, and some comparisons with multi-frequency interferometric observations towards nine Class 0/I objects. 
For our convenience in the following discussion, we define $2F^{E-bg}_{\nu}\equiv\Omega_{bg}(1-e^{-\tau^{E-bg}_{\nu}})B_{\nu}(T_{bg})e^{-\tau^{E-fg}_{\nu}}$, $2F^{E-fg}_{\nu}\equiv\Omega_{fg}(1-e^{-\tau^{E-fg}_\nu})B_{\nu}(T_{fg})$, and define $F^{B-bg}$ and $F^{B-fg}$ in similar ways.
When foreground is very optically thin, Equation \ref{eqn:multicomponent} and be simplified as

\begin{equation}\label{eqn:multicomponentthin}
   \begin{split}
      2F^{E}_{\nu} \sim  \Omega_{bg}(1-e^{-\tau^{E-bg}_{\nu}})B_{\nu}(T_{bg}) + 
           \Omega_{fg}\tau^{E-fg}_{\nu}B_{\nu}(T_{fg}) \\
      2F^{B}_{\nu} \sim  \Omega_{bg}(1-e^{-\tau^{B-bg}_{\nu}})B_{\nu}(T_{bg}) + 
           \Omega_{fg}\tau^{B-fg}_{\nu}B_{\nu}(T_{fg}).
   \end{split}
\end{equation}

To mimic the case of the observations towards inner star-forming cores which may have radial density gradients, we assume that the background component is optically very thick at all discussed frequencies (i.e., $\tau^{E-bg}_{\nu}$ $\gg$1, $\tau^{B-bg}_{\nu}$ $\gg$1).
In this case, the intensities of the background in the E and B orientations are approximately
\[ (1-e^{-\tau^{E-bg}_{\nu}})B_{\nu}(T_{bg})$ $\sim$ $(1-e^{-\tau^{B-bg}_{\nu}})B_{\nu}(T_{bg})$ $\sim$ $B_{\nu}(T_{bg}).\]
Therefore, the emission from the background component is essentially unpolarized, and that the exact values of $\tau^{E-bg}_{\nu}$ and $\tau^{B-bg}_{\nu}$ are not relevant to either the total intensity or polarization.
We note that the optically thick background component attenuates the contribution from the cosmic microwave background.
If the optically thick background component does not exist in a realistic case study, we could equivalently consider $T_{bg}=T_{CMB}$.

Under the assumption of optically very thick background, in the case that the foreground is optically very thin (e.g., Equation \ref{eqn:multicomponentthin}), the polarized intensity is contributed nearly solely by the foreground component, while the total intensity is contributed by both foreground and background emission.
Otherwise, due to foreground attenuation, the actually observed dust emission from the background component can be polarized with a polarization percentage \[P_{bg}$ $=$ $|F^{E-bg}_{\nu} - F^{B-bg}_{\nu} | / (F^{E-bg}_{\nu} + F^{B-bg}_{\nu}),\] which is nonzero if $\tau^{E-fg}_{\nu}\neq\tau^{B-fg}_{\nu}$.
This can be due to that dust grains are not spherically symmetric, and are aligned with magnetic field, radiative field, or H$_{2}$ gas flows, which can be expressed as $\tau^{E-fg}_{\nu}>\tau^{B-fg}_{\nu}$ without losing any generality for our discussion.

For the following discussion we define 

\begin{equation}\label{eqn:alpha}
   \begin{split}
      \tau^{E-fg}_{\nu}=\tau\times(1+\alpha), \\
      \tau^{B-fg}_{\nu}=\tau\times(1-\alpha),
   \end{split}
\end{equation}
with a $0\le\alpha<1$ convention.

The polarization percentage of the emission from foreground component is thereby \[P_{fg}$ $=$ $|F^{E-fg}_{\nu} - F^{B-fg}_{\nu} | / (F^{E-fg}_{\nu} + F^{B-fg}_{\nu}),\] which is approximately \[|\tau^{E-fg}_{\nu} - \tau^{B-fg}_{\nu}| / (\tau^{E-fg}_{\nu} + \tau^{B-fg}_{\nu}) = \alpha\] when $\tau\ll1$.
Apparently, the polarization position angle of the foreground-attenuated background emission is 90$^{\circ}$ offset from that of the polarized foreground emission, since that the more the foreground is emitting in one orientation, the more the electromagnetic wave from the background  is attenuated in that orientation by foreground absorption.
The overall polarization percentage \[P=| F^{E}_{\nu}-F^{B}_{\nu} | / ( F^{E}_{\nu}+F^{B}_{\nu} )\] and the polarization position angle ($P.A.$) are determined by a competition in between the foreground and background terms.
In ideal high angular resolution observations, we may assume that locally $\Omega_{bg}\sim\Omega_{fg}$ $=$ $\Omega$.
Therefore, the key factors which affect the observed polarization properties are $\tau$, $\alpha$, $T_{bg}$, and $T_{fg}$.
However, we caution that in some interferometric imaging, one may nearly completely filter out the  rather smooth and extended foreground component, which effectively makes $\Omega_{fg}\rightarrow 0$.
For regions where the actual polarization position angle (i.e., determined by $F_{\nu}^{E}$ and $F_{\nu}^{B}$) is identical to that of the foreground (alone) emission, filtering out the foreground component will make the derived polarization position angle from observations flipped by 90$^{\circ}$ since the polarized intensity is then dominated by the foreground-attenuated background emission.
We outline the related effects in spectral line polarimetric observations in Appendix  \ref{appendix:line}.

\begin{figure}
   \vspace{-1.2cm}
   \hspace{-0.3cm}
   \includegraphics[width=9.5cm]{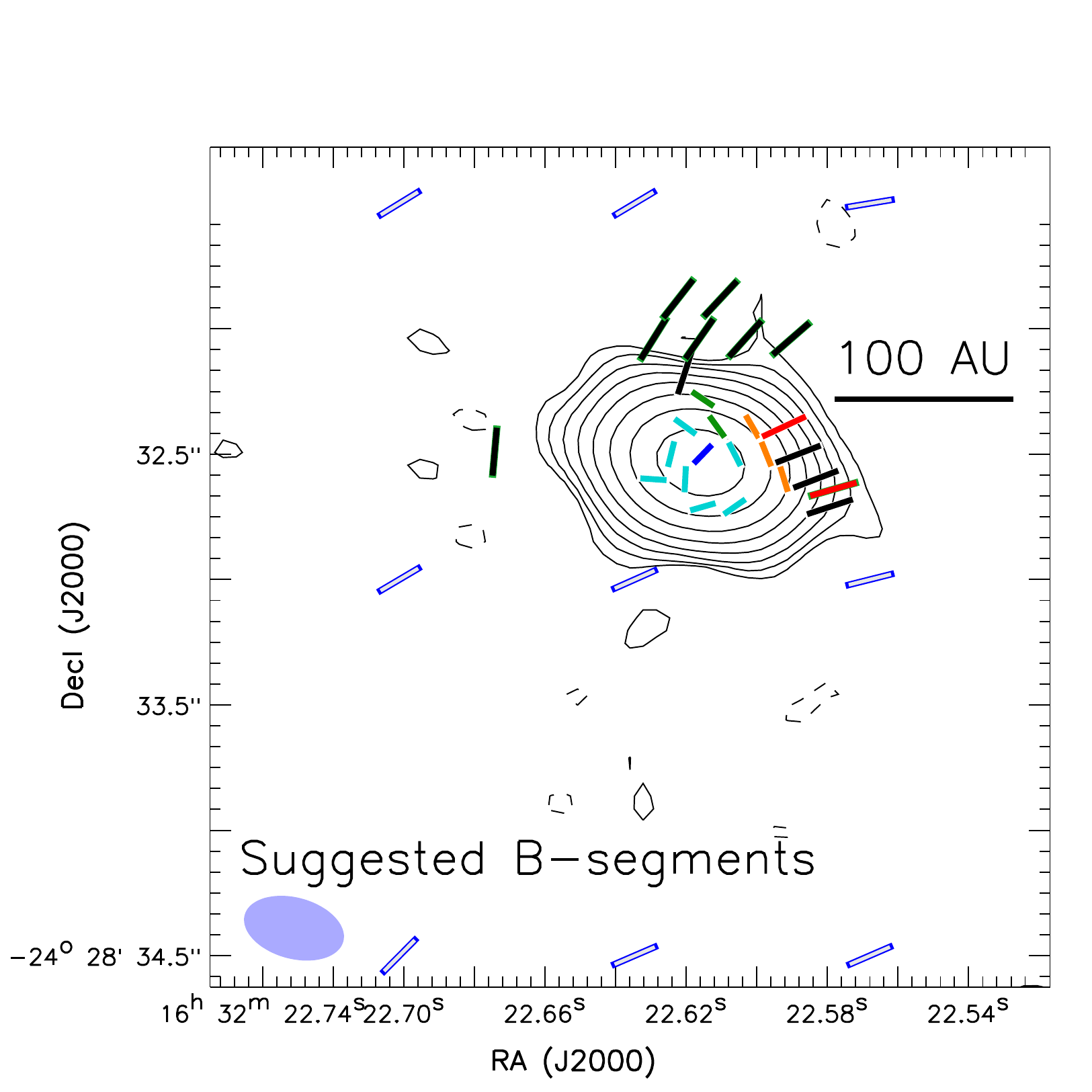}
   \caption{
      \footnotesize{
         Suggested magnetic field line segments (i.e., B-segments) around IRAS\,16293-2422\,B. Line segments bound by white lines are the inferred magnetic field orientations from the Robust = 0 weighted JVLA 40-48 GHz images, and the color coding of them is the same with those in Figure \ref{fig:jvla}; those bounded by green lines are the inferred magnetic field orientations from the Robust = 2 weighted 40-48 GHz JVLA images, those bounded by blue lines are the inferred magnetic field orientations from the Robust = 0 weighted SMA 341.5 GHz images. More detail of how this figure was generated is introduced in Section \ref{sub:interpretation}.
      }
   }
   \label{fig:Bsegments}
\end{figure}

To provide a quantitative sense about the overall polarization after considering foreground attenuation, we show in Figure \ref{fig:Permodel}, \ref{fig:Permodel2} and \ref{fig:Permodel3} the estimates of polarization percentage and position angles for $\tau=$0.1, 0.6, 1.0, 3.0, 6.0, and 10.0 at the observing frequency $\nu=$44 GHz, 345 GHz, and 690 GHz (e.g., ALMA Band 1, 7, and 9).
We limit ourselves to a generic case of $\alpha=$10\%.
The results based on other assumed $\alpha$ values are qualitatively similar, and can be easily reproduced based on our provided equations.
We show the estimates of Stokes\,I and polarization intensities in Figure \ref{fig:PImodel}, but only for $\nu=$44 GHz.
The results can be summarized as:
\begin{itemize}
   \item[1.]  When $T_{fg}>T_{bg}$, the polarization position angle is identical to that determined by the foreground emission alone. Otherwise, the polarization angle is orthogonal to that determined by the foreground emission alone. The polarization percentage increases with the  $|T_{fg}-T_{bg}|$. When $T_{fg}=T_{bg}=T$, $2F^{E}_{\nu}$ $\sim$  $\Omega B_{\nu}(T)e^{-\tau^{E-fg}_{\nu}}$ $+ \Omega(1-e^{-\tau^{E-fg}_\nu})B_{\nu}(T)$ $=$ $\Omega B_{\nu}(T)$ $=$ $2F^{B}_{\nu}$. In this case, dust continuum is essentially unpolarized. \\
   \item[2.] If $\alpha$ has no frequency dependence, then the estimated polarization percentages have weak dependence on the observing frequency $\nu$ from the $B_{\nu}(T)$ term, except in the low $T_{fg}$ or $T_{bg}$ limits such that the observations is no more in the Rayleigh-Jeans limit. Taking the $\tau=$0.1 cases for example (Figure \ref{fig:Permodel}), the foreground emission becomes more dominant when $T_{bg}$ drops below the Rayleigh-Jeans limit, which occurs earlier at $\nu=$690 GHz than 345 and 44 GHz.    
 As a result, at low $T_{bg}$ (e.g., 5-10\,K) the polarization percentages are higher at higher observing frequencies, and eventually saturate to the value of $\alpha$. For the case of $\tau\gtrsim$6 and $T_{fg}\lesssim$10\,K, the foreground component does not emit (total and polarized) flux efficiently at high frequencies, but its absorption can lead to strong polarization of the background component. Therefore, when $T_{fg}\lesssim$10\,K, we see that the polarization percentage increases significantly with observing frequency.
Observing optically thin regions at multiple frequencies which are in the Rayleigh-Jeans limit of the target sources may probe the frequency dependence of $\alpha$, which is related to the shapes of the dust grains and the grain alignment efficiency. \\
   \item[3.] For regions where $T_{bg}<T_{fg}$, the highest observable polarization percentage is $\alpha$, which is achieved when $\tau\sim1$. When $\tau$ is larger, the polarization percentage starts decreasing with $\tau$ since the intensities at the orthogonal orientations are both saturating toward the black body intensity. \\
   \item[4.] For regions where $T_{bg}>T_{fg}$, the polarization percentage can be $\gtrsim$50\%, even when $\alpha$ is only 10\%. The foreground attenuation appears to be a very efficient polarization mechanism when the difference between $T_{fg}$ and $T_{bg}$ is large. This effect is already prominent when the foreground is not completely optically thick (e.g., $\tau\sim0.6$). The polarization percentage due to foreground attenuation increases when the total intensity is becoming dominated by the lower temperature foreground, while the polarized intensity is becoming  dominated by the foreground-attenuated, high temperature background (see Figure \ref{fig:PImodel}). The polarization percentage starts decreasing with $\tau$ when the foreground is becoming so optically thick, such that the total intensity saturates to the brightness temperature of the foreground the polarized intensity from the background terms saturates to $\sim0$.
\end{itemize}

The dust polarization in fully or marginally optically thick regime outlined in this section complicates the interpretation for observations on target sources which have non-uniform density, temperature, or magnetic field direction.
Taking the specific case of a circumstellar disk in which magnetic field is perfectly toroidal and dust is aligned predominantly with magnetic field as an example, when dust emission is optically thin, in the face-on projection, the observed E-field position angles may appear either perfectly toroidal or poloidal depending on the optical depth and temperature structure in the line-of-sight direction.
In the edge-on projection (e.g., Segura-Cox et al. 2015), when the disk has a radially decreasing dust temperature profile, the observed E-field position angle will appear perpendicular to the disk rotational axis in the optically thick regime, and appear parallel to the disk rotational axis in the optically thin regime.

Polarization because of polarized foreground attenuation may to some extent explain the observed uncertain alignment in between the polarization position angle and the disk or outflow axes (e.g., Hull et al. 2014, 2017a; Galametz et al. 2018), and for some cases explain the distinct polarization components (i.e., parallel or perpendicular to the long axes of gas structures) observed around star-forming cores (e.g., Tang et al. 2010; Chen et al. 2012; Koch et al. 2014; Zhang et al. 2014; Li et al. 2015; Cortes et al. 2016; Ching et al. 2017; Pattle et al. 2017; Ward-Thompson et al. 2017; Monsch et al. submitted).
When missing Stokes I flux is not a serious concern, polarization by dust absorption may be the most natural explanation when the observed polarization percentage appears anomalously high.
Generically, we may put an operational definition of `anomalously high polarization' by that the observed polarization percentages appear considerably higher than those detected from the lower frequency, optically thinner observations towards the same region.
For regions with very high optical depths (e.g., $\tau>$3), higher than $\sim$1\% polarization percentage may favor polarization due to foreground attenuation (Figure \ref{fig:Permodel3}).
We hypothesize that the foreground-attenuation polarization is very prominent for observations at $\gtrsim$345 GHz towards hot cores which are embedded in low-temperature infrared dark clouds.
In these cases, the contrast in between $T_{bf}$ and $T_{fg}$ is large.
For detailed case studies of such sources, synthetic observations generated from hydrodynamics simulations based on an isothermal assumption (e.g., Kataoka et al. 2012; Hull et al. 2017b; Mocz et al. 2017) may compare better with observations in the fully optically thin limits.
Finally, to avoid the ambiguity in the interpretation of the observed polarization position angle, one can either observe in the fully optically thin regime, or observe sources which $T_{bg}\gg T_{fg}$, for example, against a bright background ultra compact H\textsc{ii} region.
These suggested observations may be routinely carried out with ALMA, once the Band 1 (30-50 GHz) receivers are equipped.

We remark that the foreground-absorption induced polarization is not only limited to the cases where the background dust component is optically very thick.
Setting the optical depth of the background component to be $\gg$1 is merely to simplify the considerations in our toy model (Equation \ref{eqn:multicomponent}), such that the polarization percentage of the background component can be safely ignored.

\subsection{Tentative interpretation for IRAS\,16293-2422\,B}\label{sub:interpretation}
We cannot caution more about the uncertainty of our present interpretation, given that the density and temperature structures of our target sources are not yet constrained detailed enough on our resolved spatial scales.
In addition, our 44 GHz observations are yet subject to missing short-spacing, which nevertheless, should not impact regions with high intensity.
To provide a quantitative sense of how missing short-spacing biased the observed Stokes I intensity fractionally, we assume that exterior to the disks (or disk-like structures) around IRAS\,16293-2422\,A and B, there is $\sim$2 $M_{\odot}$ of gas on $\gtrsim$10$''$ scales (Mundy et al. 1990), and assume that the gas to dust mass ratio is 100.
The averaged dust mass column density is $\Sigma_{\mbox{\tiny dust}}\sim$0.1 g\,cm$^{-2}$.
Assuming the dust opacity at 230 GHz is $\kappa_{\mbox{\tiny 230 GHz}}=$0.5$^{+0.5}_{-0.4}$ cm$^{2}$\,g$^{-1}$ (c.f., Draine et al. 2006), and the dust opacity spectral index $\beta\sim$1-1.75 (Li et al. 2017), the derived averaged dust optical depth at 44 GHz is $\tau_{\mbox{\tiny 44 GHz}}=\kappa_{\mbox{\tiny 230 GHz}}(44/230)^{\beta}\Sigma_{\mbox{\tiny dust}}=$2.8$^{+16}_{-2.2}\times$10$^{-3}$.
Assuming a nominal 20-30 K dust temperature on 10$''$ scales (van Dishoeck et al. 1995; Ceccarelli et al. 2000; Chandler et al. 2005), the brightness temperature of the extended dust emission is $\sim(25\pm5)\times(1 - e^{-\tau_{\mbox{\tiny 44 GHz}}})=$0.069$^{+0.57}_{-0.058}$ K.
This is approximately how much missing short-spacing can bias the Stokes\,I intensity.
Taking regions where the observed brightness temperature in the Robust = 0 weighted 44 GHz image is above 1.4 K (i.e., 6-$\sigma$; the second contour from bottom in the right panel of Figure \ref{fig:jvla}) for example, missing flux can bias Stokes I intensity by at most $\sim$(0.64 K)/(1.4 K + 0.64 K)$\sim$31\%.
For structures detected at the 6-$\sigma$ significance, {\tt clean} errors and thermal noise (e.g., 3-$\sigma$) may dominate the errors in the measured polarization percentages.
The brightness temperature of the missing flux can be comparable with the 6-$\sigma$ detection limit of our Robust = 2 weighted 44 GHz image ($\sim$0.6 K), and therefore the observed polarization percentages can be biased by over $\sim$50\% in the low intensity (3-6-$\sigma$) area in that image (Table \ref{tab:pol_rob2}).
Our assessment is that for our present 44 GHz images, missing short-spacing biased the observed polarization percentages of the detected, spatially rather extended dust emission component.
Since the brightness temperature of the extended component is low (Figure \ref{fig:jvla}) such that dust is likely to be optically thin (i.e., $T_{bg}=T_{CMB}$), missing short-spacing does not dramatically flip the observed polarization position angles by 90$^{\circ}$ (Section \ref{sub:mechanism}; Figures \ref{fig:Permodel}-\ref{fig:Permodel3}). 

We base on the hypothesis that the system within the inner $\sim$100 au radius around IRAS\,16293-2422\,B is an approximately face-on rotating accretion flow, which may be supported by the observed low velocity gradient across this area (e.g., Pineda et al. 2012).
We do not know the geometric scale height of the system, but expect that it is geometrically relatively thick due to a high gas temperature.
Gas flows on the larger spatial scales are collapsing towards this 100 au scale rotating system, which is supported by the detection of redshifted absorption features against the dust continuum emission (Zapata et al. 2013).
There may be an overall temperature gradient, such that the higher excitation molecular lines are preferentially detected towards the central region (e.g., van Dishoeck et al. 1995; Ceccarelli et al. 2000; Chandler et al. 2005; Oya et al. 2016; Murillo 2017; Oya et al. 2018; and see Young et al. 2018 and references therein for the more general theoretical models).
The surface layer of the rotational accretion flow may have lower temperature than the mid plane in general.
However, at the approximately central one synthesized beam area, dust at the surface layer may be significantly heated by the irradiation from host protostar, such that the detected peak brightness temperature by the 690 GHz ALMA observations is higher than that detected from our 44 GHz observations, which have comparable angular resolutions.

We assume that for all resolved spatial scales in our 44 GHz observations, the dominant polarization mechanism for the 44 GHz continuum is by the aligned dust grain.
We assume that on $>$100 au scales, the dominant polarization mechanism for the 341.5 GHz continuum is by the aligned dust grain with magnetic field.
We derive the speculated projected axis of grain alignment (i.e., the projected long axis of dust grains) based on the following assumptions:

\begin{itemize}
   \item[1.] We quote the $\sim$3$''$ angular resolution, 341.5 GHz observations from Table 4 of Rao et al. (2009), and assume that these measurements are mostly in the optically thin limit. The optically thin assumption is consistent with our estimates of the averaged dust mass column density $\Sigma_{\mbox{\tiny dust}}\sim$0.1 g\,cm$^{-2}$, assuming that the dust opacity at 345.1 GHz is $\sim$1 cm$^{2}$\,g$^{-1}$ (Ossenkopf \& Henning 1994). In this case, the observed polarization position angle is parallel to the long axis of aligned grains. \\
   \item[2.] For the Robust = 0 weighted JVLA images, we assume a nominal $\sim$30\,K average temperature of the system, and define regions with brightness temperature lower than 30\,K $\times (1 - e^{-1})$ to be in optically thin limit. Similar to case 1, the observed polarization position angle is parallel to the long axis of aligned grains. \\
   \item[3.] For the Robust = 0 weighted JVLA images, at regions where the observed brightness temperature is higher than 30\,K $\times (1 - e^{-1})$ but lower than 80\% of the peak brightness temperature, we consider that the polarization is dominated by the polarized attenuation of a cooler foreground layer. The emission from the optically thicker, warmer background component is considered to be essentially unpolarized. In this case, the observed polarization position angle is perpendicular to the long axis of aligned grains in the cooler foreground layer. \\
   \item[4.] For the Robust = 0 weighted JVLA images, for regions where the observed brightness temperature is higher than 80\% of the peak brightness temperature, we consider that the foreground, surface layer may be hotter than the optically thick background component. In this case, the observed polarization position angle is parallel to the long axis of aligned grains in the surface layer. \\
   \item[5.] For the Robust = 2 weighted JVLA images, which have poorer angular resolution does not constrain brightness temperature as well as the Robust = 0 weighted images, we only consider regions in optically thin limit selected based on rule 2.
\end{itemize}

Based on these assumptions, and assuming that the long axis of dust grains on the resolved spatial scales are aligned perpendicular to magnetic field lines, the inferred magnetic field orientation is given in Figure \ref{fig:Bsegments}.
It appears like a transition from a poloidal field configuration on $>$100 au scales to a toroidal field configuration on smaller spatial scales, which is consistent with the aforementioned geometric and kinematics pictures of IRAS\,16293-2422\,B.
On $\sim$1000 au scales, our proposed magnetic field configuration may be similar to the case of Class\,0 YSO, B335 (Maury et al. 2018).
This overall scenario is also similar to the magnetic field configuration suggested from the observations towards some intermediate or high-mass star-forming cores(Liu et al. 2013; Qiu et al. 2013; Ju{\'a}rez et al. 2017).
The most uncertain part is the central, approximately one synthesized beam area, where the optical depth is very high.
In this region, a small temperature perturbation in the line-of-sight direction can cause the observed polarization position angle to flip by 90$^{\circ}$, while the low polarization percentage is consistent with the simplified models presented in Section \ref{sub:mechanism} (see Figure \ref{fig:Permodel}, \ref{fig:Permodel2}).
We remark that the detected polarization percentages from the Robust = 0 weighted JVLA images are well in the ranges that can be explain by a constant value $\alpha\sim$10\%, except for two polarization line segments which may be biased by missing Stokes\,I flux since they are located at regions with low total intensity  (e.g., $\le$0.3 mJy\,beam$^{-1}$; see Table \ref{tab:pol_rob0}).
The polarization percentages derived from the previous SMA 341.5 GHz observations are in a similar range.

We emphasize that the magnetic field segments presented in Figure \ref{fig:Bsegments} are non-trivial and non-unique interpretation of the directly measured E-field segments presented in Figure \ref{fig:jvla}.
For the sake of objectivity, we stress the importance of presenting the direct measurements for similar types of observational studies.
Figure \ref{fig:Bsegments} is a rather qualitative picture about how the magnetic field configuration may change from large scales to the inner $\lesssim$200 au region.
We are fundamentally limited by the high dust column density, such that the observations of dust continuum are either fully in the Rayleigh-Jeans limit, or are in the extremely high optical depth limit.
Therefore, observationally it is extremely challenging to derive the constraint on the temperature and density profiles along the line-of-sight unambiguously.
Nevertheless, our proposed magnetic field configuration may be partly tested by the JVLA observations at the optically thinner bands (e.g., 8-18 GHz), as long as dust emission is still polarized at such a low frequency.
In general, to address the magnetic field configurations in the Class\,0 and I disks or disk-like structures unambiguously, we may require the sensitivity and angular resolution of the Next Generation Very Large Array (Isella et al. 2015).

\section{Conclusion}\label{sec:conclusion}
We have performed the JVLA 40-48 GHz continuum polarimetric observations towards the Class 0 YSO, IRAS\,16293-2422.
We obtained significant detection in linear polarizations.
The derived polarization position angles and polarization intensities are comparable to those derived from the 341.5 GHz SMA observations reported by Rao et al. (2009; 2014).
This may imply that both the 40-48 GHz observations and the 341.5 GHz observations trace the same polarization mechanisms that are operating on the same material in our line-of-sight.
In the case that the polarization mechanism is dominated by aligned dust grain with magnetic field lines, our suggested magnetic field configuration may be consistent with a poloidal field component converging from $\ge$100 au scales down to a toroidal field component within the inner 100 au area.
Our hypothesis may be tested by the observations at the lower frequency bands, which are optically thinner.

\begin{acknowledgements}
The Submillimeter Array is a joint project between the Smithsonian
Astrophysical Observatory and the Academia Sinica Institute of Astronomy
and Astrophysics, and is funded by the Smithsonian Institution and the
Academia Sinica (Ho et al. 2004).
The National Radio Astronomy Observatory is a facility of the National
Science Foundation operated under cooperative agreement by Associated
Universities, Inc.
Y.H. is currently supported by the Jet Propulsion Laboratory, California Institute of Technology, under a contract with the National Aeronautics and Space Administration.
Some ideas of HBL in the present manuscript were stimulated during a disk formation workshop organized by John Tobin, Catherine Walsh, and Daniel Harsono at Leiden in the summer of 2017.
\end{acknowledgements}


\clearpage
\appendix

\section{Polarization measurements}\label{appendix:poltable}
Tables \ref{tab:pol_rob2} and \ref{tab:pol_rob0} show the Stokes I, Q, U intensities, polarization position angle (P.A.) and polarization percentage (P) taken at the positions of the polarization segments presented in Figure \ref{fig:jvla}.
Details of the observations, data reduction, and imaging, are outlined in Section \ref{sec:observation}.

\begin{figure*}
   \hspace{-0.5cm}
   \begin{tabular}{ p{8.8cm} p{8.8cm} }
     \includegraphics[width=9.5cm]{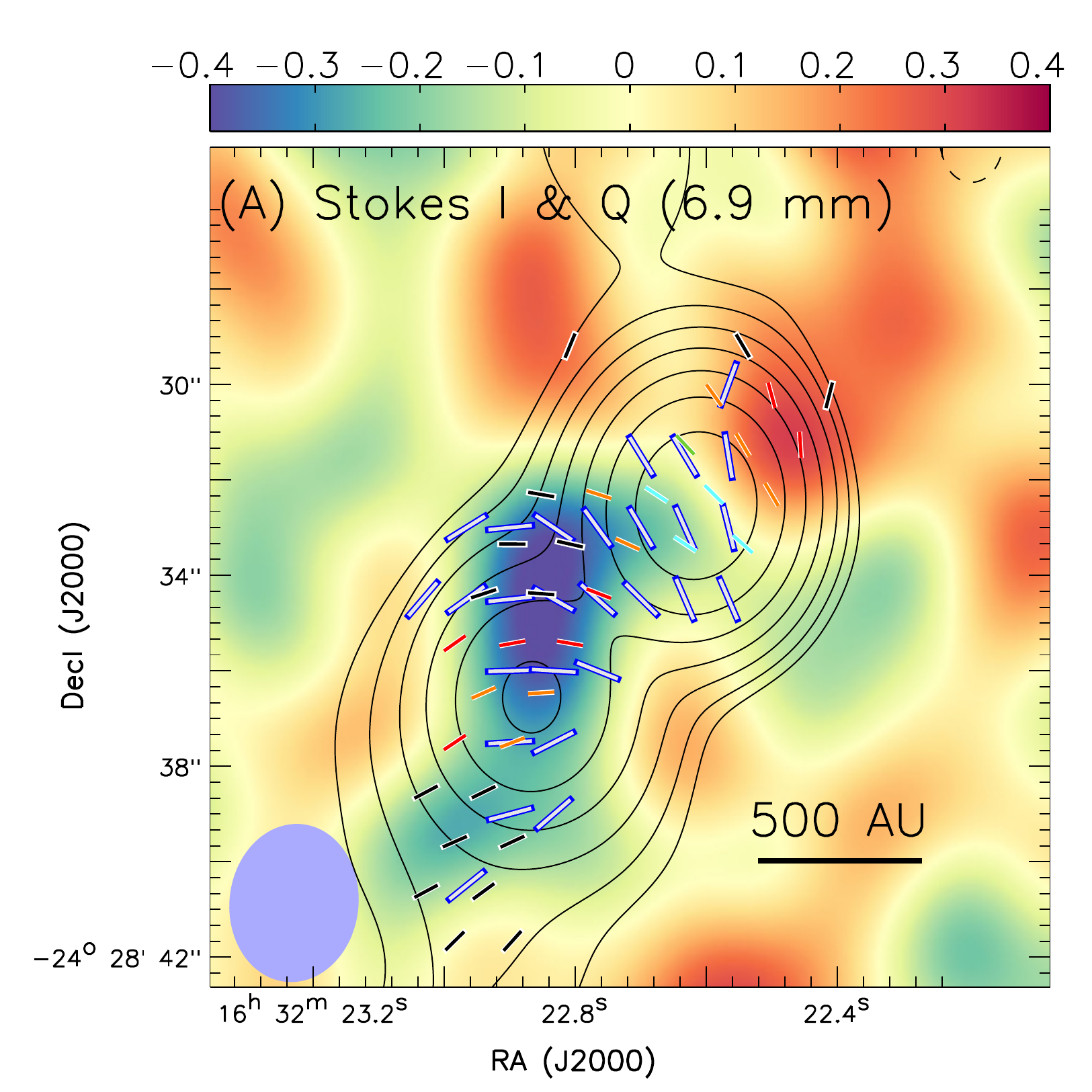} &
     \includegraphics[width=9.5cm]{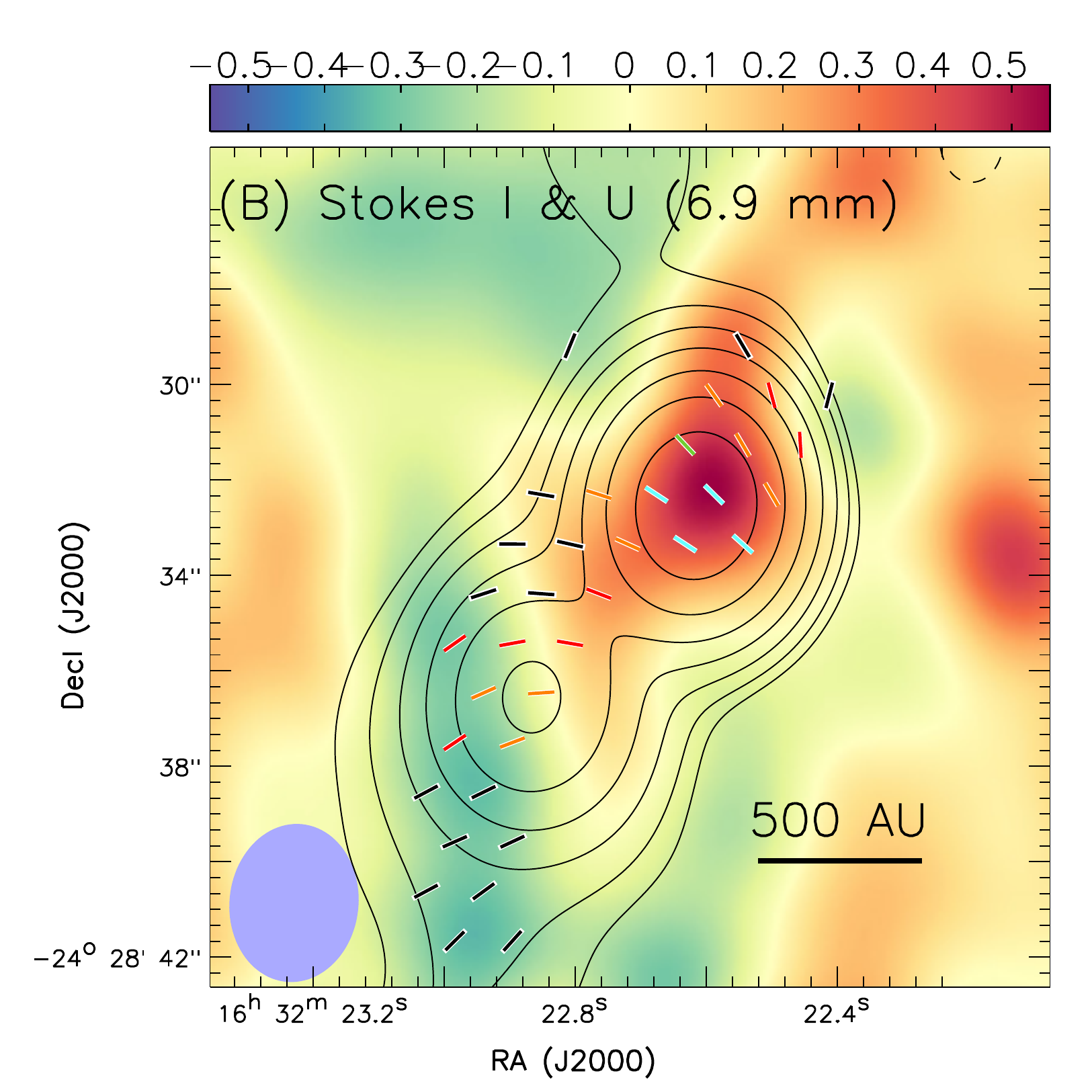} \\
   \end{tabular}
   \caption{\footnotesize{Tapered JVLA 40-48 GHz continuum images of IRAS\,16293-2422. The synthesized beam [$\theta_{\mbox{\tiny{maj}}}$   $\times$ $\theta_{\mbox{\tiny{min}}}$ $=$ 3$\farcs$3 $\times$ 2$\farcs$6 (P.A. = -5.8$^{\circ}$)] is shown in bottom left. Contours present the Stokes I intensity. The Stokes Q and U dirty (i.e., not cleaned) images are presented in colors in panel (A) and (B), respectively. Contours are 0.09 mJy\,beam$^{-1}$ (1$\sigma$, $\sim$0.0066 K) $\times$ [3, 6, 12, 24, 48, 96, 192]. Color bars are in mJy\,beam$^{-1}$ units. Line segments are similar to those presented in Figure \ref{fig:jvla}, although we caution that the polarization position angles and percentages derived from Stokes Q and U dirty images are largely uncertain. In addition, in panel (A), the SMA measurements at 341.5 GHz quoted from Table 4 of Rao et al. (2009) are overplotted as gray line segments which are bounded by blue lines. The baseline range covered by these SMA observations is 7-70\,m, which corresponds to the {\it uv} distance range of 8-80 $k\lambda$.
   }}
   \label{fig:taper}
\end{figure*}

\section{Tapered Stokes Q and U images}\label{appendix:taper}
Figure \ref{fig:taper} shows the tapered JVLA 40-48 GHz Stokes I cleaned image and Stokes Q/U dirty (i.e., not cleaned) images.
They were generated by setting the `outertaper' parameter in the CASA-{\tt clean} task to ['3arcsec', '3arcsec'].
These tapered images look qualitatively similar to the cleaned, $\sim$3$''$ resolution polarization images taken with the SMA at the central frequency of 341.5 GHz, which were reported by Rao et al. (2009).
The tapered JVLA Stokes I image is subject to significant missing flux, and therefore the imaging was dynamic range limited.
We do not have enough S/N ratios and image fidelity for robustly cleaning the JVLA Stokes Q and U images.
Therefore, we caution that the polarization position angles derived from the tapered JVLA Stokes Q and U dirty images should not be compared with other observations in a quantitative sense, and require great care even for the qualitative comparisons.
Nevertheless, the polarization position angles derived from these tapered JVLA images show similarity with those derived from the 341.5 GHz SMA observations.

We note that our JVLA observations cover the shortest {\it uv} distance of $\sim$11 $k\lambda$, which is slightly longer than that of the compared SMA observations ($\sim$8 $k\lambda$).
This may result in some difference in the observed polarization position angles.
However, the major uncertainty in the present comparison is the most likely dominated by synthesized beam pattern and thermal noise in the tapered JVLA (dirty) images.


\begin{table*}
   {\footnotesize
     \caption{\footnotesize{Polarization measurements from Robust = 2 weighted images [$\theta_{\mbox{\tiny{maj}}}$   $\times$ $\theta_{\mbox{\tiny{min}}}$ $=$ 0$\farcs$49 $\times$ 0$\farcs$38 (P.A. = 66$^{\circ}$)]}}
     \label{tab:pol_rob2}
     \begin{tabular}{cccccccc}\hline\hline
R.A. & Decl. & Stokes I & Stokes Q & Stokes U & PI & P.A. & P \\
(J2000) & (J2000) & (mJy\,beam$^{-1}$) & (mJy\,beam$^{-1}$) & (mJy\,beam$^{-1}$) & (mJy\,beam$^{-1}$) & ($^{\circ}$) & (\%) \\\hline
16$^{\mbox{\tiny{h}}}$32$^{\mbox{\tiny{m}}}$22.923$^{\mbox{\tiny{s}}}$  & -24$^{\circ}$28$'$35.67$''$ & 0.16 & -0.077 & -0.047 & 0.09 & -74 & 53 \\  
16$^{\mbox{\tiny{h}}}$32$^{\mbox{\tiny{m}}}$22.910$^{\mbox{\tiny{s}}}$  & -24$^{\circ}$28$'$35.37$''$ & 0.12 & -0.067 & -0.075 & 0.10 & -66 & 82 \\ 
16$^{\mbox{\tiny{h}}}$32$^{\mbox{\tiny{m}}}$22.910$^{\mbox{\tiny{s}}}$  & -24$^{\circ}$28$'$35.07$''$ & 0.12 & -0.039 & -0.093 & 0.10 & -56 & 77 \\ 
16$^{\mbox{\tiny{h}}}$32$^{\mbox{\tiny{m}}}$22.839$^{\mbox{\tiny{s}}}$  & -24$^{\circ}$28$'$35.82$''$ & 0.17 & -0.09 & -0.00054 & 0.09 & -90 & 49 \\ 
16$^{\mbox{\tiny{h}}}$32$^{\mbox{\tiny{m}}}$22.846$^{\mbox{\tiny{s}}}$  & -24$^{\circ}$28$'$35.67$''$ & 0.16 & -0.093 & 0.035 & 0.099 & 80 & 58 \\ 
16$^{\mbox{\tiny{h}}}$32$^{\mbox{\tiny{m}}}$22.846$^{\mbox{\tiny{s}}}$  & -24$^{\circ}$28$'$32.64$''$ & 0.11 & -0.083 & 0.05 & 0.097 & 75 & 81 \\  
16$^{\mbox{\tiny{h}}}$32$^{\mbox{\tiny{m}}}$22.839$^{\mbox{\tiny{s}}}$  & -24$^{\circ}$28$'$32.49$''$ & 0.098 & -0.092 & 0.032 & 0.097 & 80 & 95 \\ 
16$^{\mbox{\tiny{h}}}$32$^{\mbox{\tiny{m}}}$22.833$^{\mbox{\tiny{s}}}$  & -24$^{\circ}$28$'$35.37$''$ & 0.1 & -0.13 & -0.039 & 0.14 & -82 & 1.3e+02 \\ 
16$^{\mbox{\tiny{h}}}$32$^{\mbox{\tiny{m}}}$22.673$^{\mbox{\tiny{s}}}$  & -24$^{\circ}$28$'$32.49$''$ & 0.11 & -0.11 & 0.017 & 0.11 & 85 & 93 \\ 
16$^{\mbox{\tiny{h}}}$32$^{\mbox{\tiny{m}}}$22.641$^{\mbox{\tiny{s}}}$  & -24$^{\circ}$28$'$32.64$''$ & 4.2 & -0.095 & -0.005 & 0.095 & -88 & 2.1 \\ 
16$^{\mbox{\tiny{h}}}$32$^{\mbox{\tiny{m}}}$22.634$^{\mbox{\tiny{s}}}$  & -24$^{\circ}$28$'$32.49$''$ & 12 & -0.16 & 0.05 & 0.17 & 81 & 1.4 \\ 
16$^{\mbox{\tiny{h}}}$32$^{\mbox{\tiny{m}}}$22.621$^{\mbox{\tiny{s}}}$  & -24$^{\circ}$28$'$32.79$''$ & 8.8 & -0.13 & 0.02 & 0.13 & 85 & 1.4 \\  
16$^{\mbox{\tiny{h}}}$32$^{\mbox{\tiny{m}}}$22.628$^{\mbox{\tiny{s}}}$  & -24$^{\circ}$28$'$32.64$''$ & 15 & -0.21 & 0.047 & 0.22 & 84 & 1.4 \\ 
16$^{\mbox{\tiny{h}}}$32$^{\mbox{\tiny{m}}}$22.621$^{\mbox{\tiny{s}}}$  & -24$^{\circ}$28$'$32.49$''$ & 24 & -0.18 & 0.16 & 0.24 & 69 & 1 \\  
16$^{\mbox{\tiny{h}}}$32$^{\mbox{\tiny{m}}}$22.628$^{\mbox{\tiny{s}}}$  & -24$^{\circ}$28$'$32.34$''$ & 13 & -0.11 & 0.11 & 0.16 & 68 & 1.2 \\ 
16$^{\mbox{\tiny{h}}}$32$^{\mbox{\tiny{m}}}$22.621$^{\mbox{\tiny{s}}}$  & -24$^{\circ}$28$'$32.19$''$ & 5.1 & -0.098 & 0.15 & 0.18 & 61 & 3.5 \\  
16$^{\mbox{\tiny{h}}}$32$^{\mbox{\tiny{m}}}$22.628$^{\mbox{\tiny{s}}}$  & -24$^{\circ}$28$'$32.04$''$ & 0.88 & -0.051 & 0.11 & 0.12 & 58 & 13 \\ 
16$^{\mbox{\tiny{h}}}$32$^{\mbox{\tiny{m}}}$22.621$^{\mbox{\tiny{s}}}$  & -24$^{\circ}$28$'$31.88$''$ & 0.13 & -0.025 & 0.11 & 0.11 & 52 & 77 \\ 
16$^{\mbox{\tiny{h}}}$32$^{\mbox{\tiny{m}}}$22.615$^{\mbox{\tiny{s}}}$  & -24$^{\circ}$28$'$32.64$''$ & 21 & -0.19 & 0.056 & 0.20 & 82 & 0.93 \\ 
16$^{\mbox{\tiny{h}}}$32$^{\mbox{\tiny{m}}}$22.609$^{\mbox{\tiny{s}}}$  & -24$^{\circ}$28$'$32.49$''$ & 23 & -0.0059 & 0.22 & 0.22 & 46 & 0.95 \\ 
16$^{\mbox{\tiny{h}}}$32$^{\mbox{\tiny{m}}}$22.615$^{\mbox{\tiny{s}}}$  & -24$^{\circ}$28$'$32.34$''$ & 15 & -0.069 & 0.21 & 0.22 & 54 & 1.4 \\  
16$^{\mbox{\tiny{h}}}$32$^{\mbox{\tiny{m}}}$22.609$^{\mbox{\tiny{s}}}$  & -24$^{\circ}$28$'$32.19$''$ & 4.1 & -0.054 & 0.16 & 0.17 & 54 & 4 \\ 
16$^{\mbox{\tiny{h}}}$32$^{\mbox{\tiny{m}}}$22.615$^{\mbox{\tiny{s}}}$  & -24$^{\circ}$28$'$32.04$''$ & 0.77 & -0.054 & 0.15 & 0.16 & 55 & 20 \\ 
16$^{\mbox{\tiny{h}}}$32$^{\mbox{\tiny{m}}}$22.609$^{\mbox{\tiny{s}}}$  & -24$^{\circ}$28$'$31.88$''$ & 0.14 & -0.0079 & 0.11 & 0.11 & 47 & 77 \\ 
16$^{\mbox{\tiny{h}}}$32$^{\mbox{\tiny{m}}}$22.596$^{\mbox{\tiny{s}}}$  & -24$^{\circ}$28$'$32.49$''$ & 10 & 0.16 & 0.18 & 0.24 & 25 & 2.4 \\  
16$^{\mbox{\tiny{h}}}$32$^{\mbox{\tiny{m}}}$22.602$^{\mbox{\tiny{s}}}$  & -24$^{\circ}$28$'$32.34$''$ & 9.4 & 0.06 & 0.21 & 0.22 & 37 & 2.2 \\ 
16$^{\mbox{\tiny{h}}}$32$^{\mbox{\tiny{m}}}$22.602$^{\mbox{\tiny{s}}}$  & -24$^{\circ}$28$'$32.04$''$ & 0.55 & -0.014 & 0.13 & 0.13 & 48 & 23 \\  
16$^{\mbox{\tiny{h}}}$32$^{\mbox{\tiny{m}}}$22.589$^{\mbox{\tiny{s}}}$  & -24$^{\circ}$28$'$32.64$''$ & 7.4 & 0.11 & 0.076 & 0.13 & 17 & 1.7 \\ 
16$^{\mbox{\tiny{h}}}$32$^{\mbox{\tiny{m}}}$22.583$^{\mbox{\tiny{s}}}$  & -24$^{\circ}$28$'$32.49$''$ & 3.1 & 0.14 & 0.096 & 0.17 & 17 & 5.4 \\  
16$^{\mbox{\tiny{h}}}$32$^{\mbox{\tiny{m}}}$22.589$^{\mbox{\tiny{s}}}$  & -24$^{\circ}$28$'$32.34$''$ & 3.7 & 0.087 & 0.11 & 0.14 & 26 & 3.7 \\ 
16$^{\mbox{\tiny{h}}}$32$^{\mbox{\tiny{m}}}$22.589$^{\mbox{\tiny{s}}}$  & -24$^{\circ}$28$'$32.04$''$ & 0.29 & 0.012 & 0.09 & 0.09 & 41 & 29 \\ 
16$^{\mbox{\tiny{h}}}$32$^{\mbox{\tiny{m}}}$22.577$^{\mbox{\tiny{s}}}$  & -24$^{\circ}$28$'$32.64$''$ & 1.4 & 0.1 & 0.061 & 0.12 & 16 & 8.1 \\\hline
     \end{tabular}
   }
\end{table*}

\begin{table*}
   {\footnotesize
     \caption{\footnotesize{Polarization measurements from Robust = 0 weighted images [$\theta_{\mbox{\tiny{maj}}}$   $\times$ $\theta_{\mbox{\tiny{min}}}$ $=$ 0$\farcs$39 $\times$ 0$\farcs$24 (P.A. = 74$^{\circ}$)]}}
     \label{tab:pol_rob0}
     \begin{tabular}{cccccccc}\hline\hline
R.A. & Decl. & Stokes I & Stokes Q & Stokes U & PI & P.A. & P \\
(J2000) & (J2000) & (mJy\,beam$^{-1}$) & (mJy\,beam$^{-1}$) & (mJy\,beam$^{-1}$) & (mJy\,beam$^{-1}$) & ($^{\circ}$) & (\%) \\\hline
16$^{\mbox{\tiny{h}}}$32$^{\mbox{\tiny{m}}}$22.628$^{\mbox{\tiny{s}}}$  & -24$^{\circ}$28$'$32.60$''$ & 11 & -0.18 & 0.023 & 0.18 & 86 & 1.6 \\  
16$^{\mbox{\tiny{h}}}$32$^{\mbox{\tiny{m}}}$22.623$^{\mbox{\tiny{s}}}$  & -24$^{\circ}$28$'$32.50$''$ & 17 & -0.18 & 0.091 & 0.20 & 76 & 1.2 \\  
16$^{\mbox{\tiny{h}}}$32$^{\mbox{\tiny{m}}}$22.614$^{\mbox{\tiny{s}}}$  & -24$^{\circ}$28$'$32.71$''$ & 11 & -0.11 & -0.072 & 0.13 & -74 & 1.2 \\  
16$^{\mbox{\tiny{h}}}$32$^{\mbox{\tiny{m}}}$22.619$^{\mbox{\tiny{s}}}$  & -24$^{\circ}$28$'$32.60$''$ & 17 & -0.22 & 0.027 & 0.22 & 87 & 1.3 \\  
16$^{\mbox{\tiny{h}}}$32$^{\mbox{\tiny{m}}}$22.614$^{\mbox{\tiny{s}}}$  & -24$^{\circ}$28$'$32.50$''$ & 20 & -0.0043 & 0.18 & 0.18 & 46 & 0.9 \\  
16$^{\mbox{\tiny{h}}}$32$^{\mbox{\tiny{m}}}$22.619$^{\mbox{\tiny{s}}}$  & -24$^{\circ}$28$'$32.39$''$ & 12 & -0.047 & 0.15 & 0.16 & 54 & 1.2 \\  
16$^{\mbox{\tiny{h}}}$32$^{\mbox{\tiny{m}}}$22.614$^{\mbox{\tiny{s}}}$  & -24$^{\circ}$28$'$32.28$''$ & 4.3 & -0.044 & 0.11 & 0.12 & 56 & 2.7 \\  
16$^{\mbox{\tiny{h}}}$32$^{\mbox{\tiny{m}}}$22.619$^{\mbox{\tiny{s}}}$  & -24$^{\circ}$28$'$32.17$''$ & 0.25 & -0.095 & 0.067 & 0.12 & 72 & 45 \\  
16$^{\mbox{\tiny{h}}}$32$^{\mbox{\tiny{m}}}$22.605$^{\mbox{\tiny{s}}}$  & -24$^{\circ}$28$'$32.71$''$ & 9.1 & -0.044 & -0.12 & 0.13 & -55 & 1.3 \\  
16$^{\mbox{\tiny{h}}}$32$^{\mbox{\tiny{m}}}$22.605$^{\mbox{\tiny{s}}}$  & -24$^{\circ}$28$'$32.50$''$ & 15 & 0.15 & 0.21 & 0.26 & 27 & 1.7 \\ 
16$^{\mbox{\tiny{h}}}$32$^{\mbox{\tiny{m}}}$22.610$^{\mbox{\tiny{s}}}$  & -24$^{\circ}$28$'$32.39$''$ & 11 & 0.079 & 0.24 & 0.25 & 36 & 2.4 \\ 
16$^{\mbox{\tiny{h}}}$32$^{\mbox{\tiny{m}}}$22.596$^{\mbox{\tiny{s}}}$  & -24$^{\circ}$28$'$32.50$''$ & 5.2 & 0.18 & 0.18 & 0.25 & 22 & 4.9 \\  
16$^{\mbox{\tiny{h}}}$32$^{\mbox{\tiny{m}}}$22.600$^{\mbox{\tiny{s}}}$  & -24$^{\circ}$28$'$32.39$''$ & 6.4 & 0.12 & 0.21 & 0.24 & 30 & 3.8 \\  
16$^{\mbox{\tiny{h}}}$32$^{\mbox{\tiny{m}}}$22.591$^{\mbox{\tiny{s}}}$  & -24$^{\circ}$28$'$32.60$''$ & 3.6 & 0.14 & 0.09 & 0.17 & 17 & 4.4 \\  
16$^{\mbox{\tiny{h}}}$32$^{\mbox{\tiny{m}}}$22.587$^{\mbox{\tiny{s}}}$  & -24$^{\circ}$28$'$32.50$''$ & 1.7 & 0.13 & 0.12 & 0.18 & 21 & 10 \\ 
16$^{\mbox{\tiny{h}}}$32$^{\mbox{\tiny{m}}}$22.591$^{\mbox{\tiny{s}}}$  & -24$^{\circ}$28$'$32.39$''$ & 1.6 & 0.076 & 0.093 & 0.12 & 25 & 7.3 \\ 
16$^{\mbox{\tiny{h}}}$32$^{\mbox{\tiny{m}}}$22.578$^{\mbox{\tiny{s}}}$  & -24$^{\circ}$28$'$32.71$''$ & 0.3 & 0.097 & 0.07 & 0.12 & 18 & 38 \\ 
16$^{\mbox{\tiny{h}}}$32$^{\mbox{\tiny{m}}}$22.582$^{\mbox{\tiny{s}}}$  & -24$^{\circ}$28$'$32.60$''$ & 1.2 & 0.12 & 0.11 & 0.16 & 21 & 14 \\ \hline
     \end{tabular}
   }
\end{table*}

\section{Anomalous spectral line polarization due to continuum subtraction or missing short-spacing}\label{appendix:line}
In this section we base on a toy model to introduce how in the optically thick regime, dust continuum subtraction, and missing short-spacing (e.g., in interferometric observations) may anomalously polarize the observed spectral lines.
We use the terms defined for Equation \ref{eqn:multicomponent}, and additionally include a spectral line (e.g., $^{12}$CO) emission component which is foreground to all dust components assumed for Equation \ref{eqn:multicomponent}.
We refer to Girart et al. (1999), Lai et al. (2003), and Ching et al. (2016) as some likely relevant observational case studies.

For spectroscopic observations of which the observing frequency (or velocity) is offset from that of the foreground line emission, the observed fluxes in the locally defined, orthogonal E and B orientations can be expressed as
\begin{equation}\label{eqn:line_off}
\begin{split}
F_{\nu}^{E-off} =  (F_{\nu}^{E-bg} + F_{\nu}^{E-fg}) \\
F_{\nu}^{B-off} =  (F_{\nu}^{B-bg} + F_{\nu}^{B-fg}), \\
\end{split}
\end{equation}

otherwise,

\begin{equation}\label{eqn:line_on}
\begin{split}
F_{\nu}^{E-on} =  (F_{\nu}^{E-bg} + F_{\nu}^{E-fg})e^{-\tau_{\nu}^{E-mol}} +
                  \frac{1}{2}\Omega_{mol}(1 - \tau_{\nu}^{E-mol})B_{\nu}(T_{mol}) \\
F_{\nu}^{B-on} =  (F_{\nu}^{B-bg} + F_{\nu}^{B-fg})e^{-\tau_{\nu}^{B-mol}} +
                  \frac{1}{2}\Omega_{mol}(1 - \tau_{\nu}^{B-mol})B_{\nu}(T_{mol}), \\
\end{split}
\end{equation}
where $T_{mol}$ is the equivalent excitation temperature of the observed molecular line transition, $\tau_{\nu}^{E-mol}$ and $\tau_{\nu}^{B-mol}$ are the optical depths of the foreground molecular component, and $\Omega_{mol}$ is the solid angle of the foreground component,
which we assume to be identical to those of the background dust components.

When $\tau_{\nu}^{E-mol}\gg1$ and $\tau_{\nu}^{B-mol}\gg1$, approximately $F_{\nu}^{E-on}$ $\sim$ $F_{\nu}^{B-on}$ $\sim$ $\frac{1}{2}\Omega_{mol}B_{\nu}(T_{mol})$.
In all single dish (sub-)millimeter observations and most of the interferometric observations, the continuum baseline levels are subtracted in post processings.
The continuum baseline subtraction is fundamentally required for all present and previous generation single dish telescopes, for removing the atmospheric and passband features.
If the continuum subtraction is performed for individual polarizations separately, then the continuum-subtracted  spectral line fluxes are:

\begin{equation}\label{eqn:line_sub}
\begin{split}
F_{\nu}^{E-sub} \sim   -F_{\nu}^{E-off} +
                  \frac{1}{2}\Omega_{mol}B_{\nu}(T_{mol}) \\
F_{\nu}^{B-sub} \sim   -F_{\nu}^{B-off} +
                  \frac{1}{2}\Omega_{mol}B_{\nu}(T_{mol}). \\
\end{split}
\end{equation}

For optically thick regions, $F_{\nu}^{E-off, B-off}$ may be comparable with $\frac{1}{2}\Omega_{mol}B_{\nu}(T_{mol})$.
If $\frac{1}{2}\Omega_{mol}B_{\nu}(T_{mol})>F_{\nu}^{E-off, B-off}$, which is typical for the observations of optically thick line emission (e.g., $^{12}$CO), and assuming $F_{\nu}^{E-off}>F_{\nu}^{B-off}$ due to any polarization mechanisms introduced in Sectin \ref{sec:discussion} for example, then we obtain $F_{\nu}^{E-sub}<F_{\nu}^{B-sub}$.
As a result, spectral line can appear (potentially very largely) anomalously polarized after continuum subtraction, with polarization position angle 90$^{\circ}$ offset from that of the background dust continuum, even in the case that the molecular line emission is not polarized by itself.
On the other hand, in the case that $\frac{1}{2}\Omega_{mol}B_{\nu}(T_{mol})<F_{\nu}^{E-off, B-off}$, we will detect absorption line with negative intensity after performing continuum subtraction, with polarization position angle identical to that of the background emission source.
As an extreme case with $\frac{1}{2}\Omega_{mol}B_{\nu}(T_{mol})\ll F_{\nu}^{E-off, B-off}$, the C\textsc{i} line absorption at 492 GHz against the Sgr\,A* was found to have similar polarization property with the continuum emission of Sgr\,A* (Liu et al. 2016).
As a conclusion, for single-dish spectral polarimetric observations towards optically thick regions, it is necessary to jointly interpret the spectral line data with some complementary continuum polarimetric mapping observations.
When the targeted spectral line is optically very thick, it is useful to check the spectral line polarization before performing continuum baseline fitting and subtraction.

We can further consider the case that the background continuum component is spatially more extended than the observed lines.
It is general that the spectral line emission in each narrow (e.g., $\lesssim$0.3 km\,s$^{-1}$) velocity channel is spatially less extended than dust emission around star-forming cores.
In addition, we consider the case that dust grains are aligned with rather uniform, large-scale magnetic field.
In this case, we can express $F_{\nu}^{E-off, B-off}$ as $\bar{F}_{\nu}^{E-off, B-off}$ + $\hat{F}_{\nu}^{E-off, B-off}$, where the first term denotes the spatially uniform component, and the second is the local perturbation.
After considering the missing short-spacing (i.e., high pass filtering), approximately we have $F_{\nu}^{E-off, B-off}$ $\sim$ $\hat{F}_{\nu}^{E-off, B-off}$, although this also depends on exactly how the telescope (array) responds to structures of various angular scales.
The observed fluxes of optically thick line after considering missing short-spacing are approximately
\begin{equation}\label{eqn:filter_on}
\begin{split}
F_{\nu}^{E-on} =  -\bar{F}_{\nu}^{E-off} +
                  \frac{1}{2}\Omega_{mol}B_{\nu}(T_{mol}) \\
F_{\nu}^{B-on} =  -\bar{F}_{\nu}^{B-off} +
                  \frac{1}{2}\Omega_{mol}B_{\nu}(T_{mol}). \\
\end{split}
\end{equation}
After performing continuum baseline fitting and subtraction, we are left with the line fluxes

\begin{equation}\label{eqn:filter_on_sub}
\begin{split}
F_{\nu}^{E-sub} \sim   -\bar{F}_{\nu}^{E-off} - \hat{F}_{\nu}^{E-off} +
                  \frac{1}{2}\Omega_{mol}B_{\nu}(T_{mol}) \\
F_{\nu}^{B-sub} \sim   -\bar{F}_{\nu}^{B-off} - \hat{F}_{\nu}^{E-off} +
                  \frac{1}{2}\Omega_{mol}B_{\nu}(T_{mol}), \\
\end{split}
\end{equation}
which may be further approximated by 
\begin{equation}\label{eqn:filter_on_sub_approx}
\begin{split}
F_{\nu}^{E-sub} \sim   -\bar{F}_{\nu}^{E-off} +
                  \frac{1}{2}\Omega_{mol}B_{\nu}(T_{mol}) \\
F_{\nu}^{B-sub} \sim   -\bar{F}_{\nu}^{B-off} +
                  \frac{1}{2}\Omega_{mol}B_{\nu}(T_{mol}), \\
\end{split}
\end{equation}
if the local perturbation is small (i.e., $\bar{F}_{\nu}^{E-off, B-off} \gg \hat{F}_{\nu}^{E-off, B-off}$.

Observationally, in the case that continuum emission is subject to serious missing short-spacing issue, and spectal lines are spatially compact over a narrow velocity range, what we can directly and independently measure are $\hat{F}_{\nu}^{E-off, B-off}$ and $F_{\nu}^{E-sub, B-sub}$.
In other words, the spatially compact and optically thick line components can serve as small-scale masks which preserve the polarization information of the spatially extended continuum emission background to the line emission component, even in the case that the spatially extended continuum emission component is completely filtered out in the observed continuum image (or visibilities in the case of interferometric observations).
In the case that the foreground line brightness is much lower than that of the background continuum (i.e., $\frac{1}{2}\Omega_{mol}B_{\nu}(T_{mol}) \ll \bar{F}_{\nu}^{E-off, B-off})$, a comparison of the polarization position angles observed from spectral line and continuum may provide a less biased sense of how magnetic field is locally perturbed (e.g., by turbulence) than by analyzing the high-pass filtered continuum image alone, which may help derive magnetic field strength more accurately (Chandresekhar \& Fermi 1953).

The caveat is that in these derivations we assume that there is only one dominant magnetic field component in a line-of-sight.
If the magnetic field is tangled, it may require the observations of spectral line and continuum polarization at a distribution of excitation conditions and frequencies, and then study the system in a tomographic sense.
In any case, any interpretation of continuum or line polarization in the optically thick regime needs great care.

\end{document}